\documentclass[twocolumn,11pt]{aastex61}
\usepackage{aas_macros}
\usepackage{lineno}
\usepackage{latexsym}
\usepackage[utf8]{inputenc} 
\usepackage{verbatim, amsmath, amsfonts, amssymb,amssymb}
\usepackage{bigints}
\usepackage{pifont}
\usepackage{enumitem}
\usepackage{booktabs}
\usepackage{multirow}
\usepackage[T1]{fontenc}
\usepackage{empheq}

\definecolor{orangex}{HTML}{d95f02}
\definecolor{bluex}{HTML}{377eb8}
\definecolor{greenx}{HTML}{4daf4a}
\definecolor{redx}{HTML}{e41a1c}
\definecolor{purplex}{HTML}{984ea3}

\usepackage[skins,theorems]{tcolorbox}
\tcbset{highlight math style={enhanced,
 colframe=cyan!10!gray,colback=gray!5!white,arc=4pt,boxrule=1pt,
  drop fuzzy shadow}}

\newenvironment{alignteo}%
  {\empheq[box=\tcbhighmath]{align}}
  {\endempheq}

\newcommand{\qq}[3]{{#1}_{-#2}^{+#3}}
\newcommand{\pp}[2]{{#1}\pm{#2}}

\newcommand{\BeLi}{\textsuperscript{7}B\MakeLowercase{e(n,p)}\textsuperscript{7}L\MakeLowercase{i}
}
\newcommand{\LiBe}{\textsuperscript{7}L\MakeLowercase{i(p,n)}\textsuperscript{7}B\MakeLowercase{e}
}

\received{\today}
\submitjournal{ApJ}

\begin{document}

\title{Hierarchical Bayesian Thermonuclear Rate for the \BeLi Big Bang Nucleosynthesis Reaction}

\author[0000-0001-7207-4584]{Rafael S. de Souza}
\affiliation{Department of Physics \& Astronomy, University of North Carolina at Chapel Hill, NC 27599-3255, USA}
\affiliation{Triangle Universities Nuclear Laboratory (TUNL), Durham, North Carolina 27708, USA}

\author{Tan Hong Kiat}
\affiliation{National University of Singapore, 119077 Singapore}

\author{Alain Coc}
\affiliation{Centre de Sciences Nucl\'eaires et de Sciences de la Mati\`ere, Univ. Paris-Sud, CNRS/IN2P3, Universit\'e Paris-Saclay, B\^atiment, 104, F-91405 Orsay Campus, France}

\author[0000-0003-2381-0412]{Christian Iliadis}
\affiliation{Department of Physics \& Astronomy, University of North Carolina at Chapel Hill, NC 27599-3255, USA}
\affiliation{Triangle Universities Nuclear Laboratory (TUNL), Durham, North Carolina 27708, USA}

\correspondingauthor{Rafael S. de Souza}
\email{drsouza@ad.unc.edu}
\correspondingauthor{Christian Iliadis}
\email{iliadis@unc.edu}

\begin{abstract}
Big bang nucleosynthesis provides the earliest probe of standard model physics, at a time when the universe was  less than a thousand seconds old. It determines the abundances of the lightest nuclides, which give rise to the subsequent history of the visible matter in the Universe. This work derives new \BeLi thermonuclear reaction rates based on all available experimental information. This reaction sensitively impacts the primordial abundances of $^{7}$Be and $^7$Li during big bang nucleosynthesis. We critically evaluate all available data and disregard experimental results that are questionable. For the nuclear model, we adopt an incoherent sum of single-level, two-channel R-matrix approximation expressions, which are implemented into a hierarchical Bayesian model, to analyze the remaining six data sets we deem most reliable. In the fitting of the data, we consistently model all known sources of uncertainty,  including discrepant absolute normalizations of different data sets, and also take the variation of the neutron and proton channel radii into account, hence providing less biased estimates of the $^7$Be(n,p)$^7$Li thermonuclear rates. From the resulting posteriors, we extract R-matrix parameters ($E_r$, $\gamma^2_n$, $\gamma^2_p$) and derive excitation energies, partial and total widths. Our fit is sensitive to the contributions of the first three levels above the neutron threshold. Reaction rates were computed by integrating 10,000 samples of the reduced cross section. Our \BeLi thermonuclear rates have uncertainties between 1.5\% and 2.0\% at temperatures of $\leq$1~GK. We compare our rates to previously  results and find that the \BeLi rates most commonly used in big bang simulations have too optimistic uncertainties.
\end{abstract}

\keywords{methods: numerical --- nuclear reactions, nucleosynthesis, abundances --- stars: interiors --- primordial nucleosynthesis}

\section{Introduction}\label{sec:intro}
Big bang nucleosynthesis (BBN) provides the earliest probe of standard model physics, at a time when the Universe was  less than a thousand seconds old. It determines the abundances of the lightest nuclides, $^1$H, $^2$H, $^3$He, $^4$He, and $^7$Li, which give rise to the subsequent history of the visible matter in the Universe. The current uncertainties for the observed primordial abundances of $^4$He, $^2$H, and $^7$Li amount to 1.6\%, 1.2\%, and 20\%, respectively \citep{ave15,coo18,sbo10}, while for the observed primordial $^3$He abundance, only an upper limit is available ($^3$He/H $\leq$ $1.3 \times 10^{-5}$; \citealt{ban02}). To reduce the uncertainties in the predicted abundances to the level of the observational results, the rates of the most important nuclear reactions must be known to within a few percent uncertainty. At present, the uncertainties in the predicted abundances of $^4$He, $^2$H (or D), $^3$He, and $^7$Li amount to 0.07\%, 1.5\%, 2.4\%, and 4.4\%, respectively \citep{pit18}.

\begin{figure}
\includegraphics[width=\linewidth]{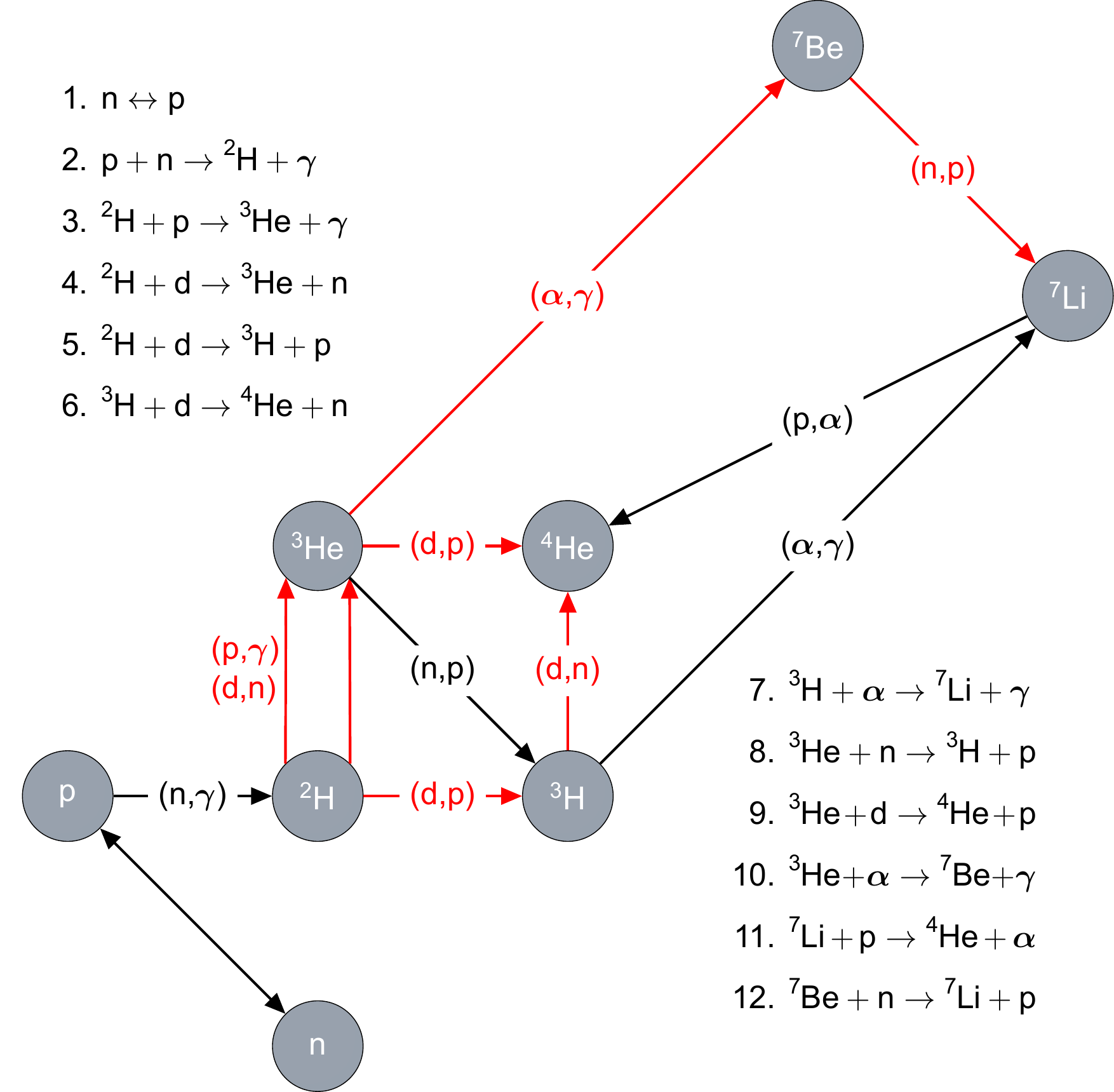}
\caption{Nuclear reactions important for big bang nucleosynthesis (BBN). Reactions for which the rates have been obtained using Bayesian models, including the present work, are shown as red arrows. 
\label{fig:BBN}}
\end{figure}

Figure~\ref{fig:BBN} shows the twelve nuclear processes of interest that take place during BBN. Among these are the weak interactions that transform neutrons into protons, and vice versa, and the p(n,$\gamma$)d reaction whose cross section can be calculated precisely using effective field theories \citep{sav99,and06}. The ten remaining reactions, $^2$H(p,$\gamma$)$^3$He, $^2$H(d,n)$^3$He, $^2$H(d,p)$^3$H, $^3$H(d,n)$^4$He, $^3$H($\alpha$,$\gamma$)$^7$Li, $^3$He(d,p)$^4$He, $^3$He(n,p)$^3$H, $^3$He($\alpha$,$\gamma$)$^7$Be, $^7$Li(p,$\alpha$)$^4$He, and $^7$Be(n,p)$^7$Li, have been measured directly in the laboratory at the energies of astrophysical interest. However, the estimation of thermonuclear reaction rates from the measured cross section data remains challenging. Previous results obtained using $\chi^2$ optimization are plagued by a number of problems, for example, the treatment of systematic uncertainties and data sets of vastly different sizes. Recently, statistically sound BBN reaction rates have been derived using hierarchical Bayesian models for the following reactions: $^2$H(p,$\gamma$)$^3$He, $^3$He($\alpha$,$\gamma$)$^7$Be \citep{ili16}, $^2$H(d,n)$^3$He, $^2$H(d,p)$^3$H \citep{gom17},  $^3$H(d,n)$^4$He \citep{des19b}, $^3$He(d,p)$^4$He \citep{des19a}.

This work reports the Bayesian reaction rates for the seventh BBN reaction, $^7$Be(n,p)$^7$Li, which sensitively influences the primordial abundance of $^7$Be. For example, a reaction rate uncertainty of $\approx$ 5\% at BBN temperatures translates to a $\approx$ 4\% variation in the predicted abundance of $^7$Li \citep{coc10}. Reliable estimations of the rates for all reactions impacting the $^7$Li or $^7$Be abundances are particularly important because the predicted $^7$Li/H ratio \citep{cyb16} exceeds the observed one \citep{sbo10} by a factor of $\approx$ $3$. This long-standing ``cosmological lithium problem'' has not found a satisfactory solution yet. A factor of $\approx$3.5 increase in the $^7$Be(n,p)$^7$Li rate would reconcile the predicted and observed primordial $^7$Li abundances. Although we did not expect such a large change at the outset of our study, a more reliable $^7$Be(n,p)$^7$Li rate is highly desirable for improving BBN predictions. 

Most of the primordial $^7$Li is produced as $^7$Be during the latter stages of BBN, when the temperature has declined to a value near $\approx$0.5~GK. This temperature corresponds to $^7$Be $+$ $n$ center-of-mass energies between $10^{-6}$~MeV to $0.3$~MeV. The $^7$Be(n,p)$^7$Li reaction near the neutron threshold has been measured by several groups, both at thermal and non-thermal neutron energies. In addition, measurements of the time-reverse $^7$Li(p,n)$^7$Be reaction provide valuable cross section information. Previous work has either used indiscriminately all available data or adopted results from arbitrary subsets of experiments for calculating the reaction rates \citep[e.g.,][]{ada03,des04,dam18}. For our analysis, we firstly present a critical evaluation of all published data, and will subsequently adopt only those experimental results that we deem to be reliable.  It will become obvious in later sections that, despite this effort, significant inconsistencies remain between the evaluated data from different measurements. It is thus interesting to devise strategies for including the various sources of statistical and systematic uncertainties into the data analysis.

The $^7$Be(n,p)$^7$Li reaction has been previously analyzed by several groups using R-matrix theory \citep[e.g.,][]{koe88,ada03,des04}. The $^7$Be(n,p)$^7$Li cross section is strongly enhanced near the neutron threshold because of a 2$^-$ level (s-wave resonance) in $^8$Be. The cross section at thermal neutron energy amounts to $\approx$ $4.5\times10^4$~barn, which is the largest thermal cross section known in the region of the light nuclides. The relative magnitudes of the partial widths for this level in the self-conjugate $^8$Be nucleus, and the implications for isospin mixing, have been debated in the literature for the past decades \citep[see, e.g.,][]{bar77,koe88}.

To fit the data using R-matrix theory (or its single-level approximation), recent work \citep[e.g.,][]{kun16,dam18} adopted the $^8$Be nuclear structure information from the Evaluated Nuclear Structure Data File (ENSDF)\footnote{From ENSDF database as of September 30, 2019. Version available at \url{http://www.nndc.bnl.gov/ensarchivals/.}} and kept the excitation energies fixed in the fitting. This procedure is problematic since several $^8$Be levels in the relevant excitation energy range are reported in ENSDF without any energy uncertainties. Even for levels with an assigned excitation energy uncertainty, the reported values are questionable, considering that all levels near the $^8$Be neutron threshold are very broad, with total widths ranging between $100$~keV and $1$~MeV, and that the excitation energies were sometimes extracted from rather featureless pulse-height spectra of the original works. 

The goal of the present work is to analyze the $^7$Be(n,p)$^7$Li cross section near the neutron threshold by incorporating expressions from the single-level, two-channel approximation of R-matrix theory into a hierarchical Bayesian model. In Section~\ref{sec:data}, we evaluate and select the data for further analysis. The reaction formalism is summarized in Section~\ref{sec:rmatrix}. Our Bayesian model, and its application to the $^7$Be(n,p)$^7$Li reaction, is discussed in Section~\ref{sec:bayes}. In Section~\ref{sec:structure}, we evaluate the properties of $^8$Be levels near the neutron threshold. Results of our Bayesian R-matrix fit are presented in Section~\ref{sec:analysis}. Thermonuclear reaction rates are given in Section~\ref{sec:rates}. A summary and conclusions are provided in Section~\ref{sec:summary}. Details about our data evaluation are discussed in Appendix \ref{sec:app}. All energies in this work are given in the center-of-mass system, unless mentioned otherwise.

\section{Data Selection and Evaluation}
\label{sec:data}
A rigorous data analysis necessitates a careful evaluation and selection of the available data. Details regarding our methods are provided in Appendix \ref{sec:app}. In brief, we started by evaluating the original data for each relevant experiment. Although used in some previous analyses \citep{ada03,des04}, we disregarded certain data sets if we had reasons for questioning their reliability or if we were unable to determine statistical and systematic uncertainties. For each adopted experiment, we also examined the experimental energy range and disregarded data points that were significantly affected by experimental artifacts. 

We also adopted the results of two $^7$Li(p,n)$^7$Be measurements \citep{gib59,mar19}, and transformed the data to $^7$Be(n,p)$^7$Li cross sections using the reciprocity theorem (see Appendix~\ref{sec:reciprocity}). We only considered data for proton laboratory energies of $E_p$ $\le$ $2371$~keV (or neutron center-of-mass energies of $E_{^7\rm{Be+n}}$ $\le$ $420$~keV), since at higher energies the neutron channel to the first excited $^7$Be state at $429$~keV is open. 

\begin{figure*}[hbt!]
\includegraphics[width=1\linewidth]{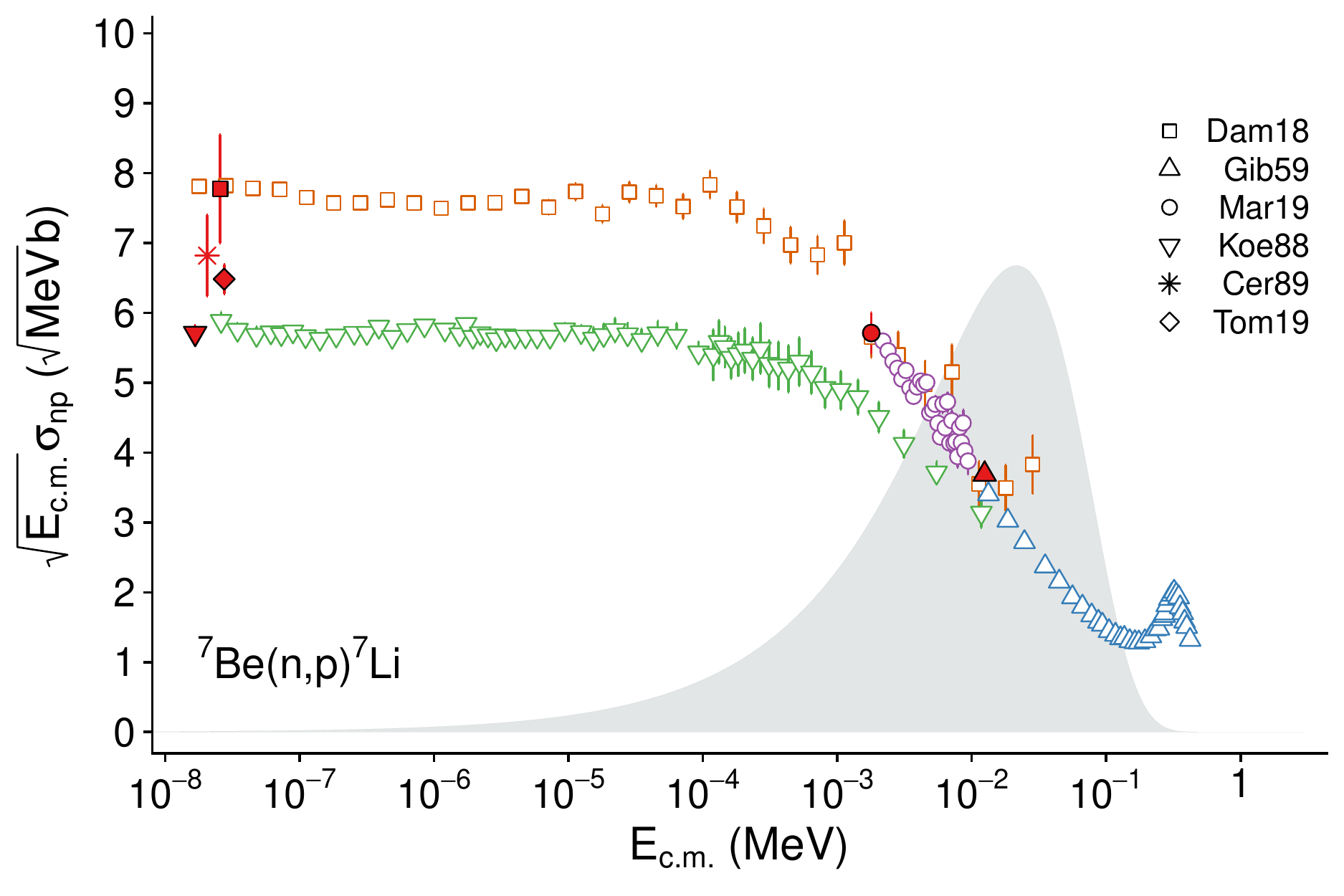}
\caption{\label{fig:data} 
The $^7$Be(n,p)$^7$Li data analyzed in the present work. The ordinate and abscissa show the reduced cross section and the center-of-mass neutron energy, respectively.  Relative cross section data are shown as open symbols: green inverted triangles \citep[\textcolor{greenx}{$\triangledown$};][]{koe88}; orange squares \citep[\textcolor{orangex}{$\square$};][]{dam18}; blue triangles \citep[\textcolor{bluex}{$\triangle$};][]{gib59}; purple circles \citep[\textcolor{purplex}{$\circ$};][]{mar19}. For these data, only statistical uncertainties are shown. Data with an absolute cross section normalization are shown as full red symbols: asterisk  \citep[\textcolor{redx}{$\ast$};][]{cer89}; square \citep[\textcolor{redx}{$\blacksquare$};][]{dam18}; diamond \citep[\textcolor{redx}{$\blacklozenge$};][]{tom19}; inverted triangle \citep[\textcolor{redx}{$\blacktriangledown$};][]{koe88}; triangle \citep[\textcolor{redx}{$\blacktriangle$};][]{gib59}, \citep[\textcolor{redx}{$\bullet$};][]{mar16}. For these data, the systematic and statistical uncertainties are shown, which have been added in quadrature. The thermal (absolute) cross sections (at the lowest energies) have been slightly shifted in energy for improved visibility. For details regarding the data selection, evaluation, and uncertainties, see Appendix \ref{sec:app}. All data shown, except those of \citet{dam18} and of \citet{koe88}, correspond to the population of the $^7$Li ground state alone. The shaded region indicates the energy range important for big bang nucleosynthesis at a temperature of $0.5$~GK.
}
\end{figure*}

The data adopted in the present analysis are shown in Figure~\ref{fig:data}. The ordinate and abscissa display the reduced (n,p) cross section, $\sqrt{E_{c.m.}}\sigma_{np}$, versus the neutron center-of-mass energy, $E_{c.m.}$.  The open symbols show relative data (i.e., those without an absolute cross section normalization; see Section~\ref{sec:bay7be}): the green inverted triangles and orange squares show the non-thermal data of \citet{koe88} and \citet{dam18}, respectively; the blue triangles and purple circles correspond to the transformed $^7$Li(p,n)$^7$Be cross sections of \citet{gib59} and \citet{mar19}, respectively. The full red data points denote those with an absolute cross section normalization (Section~\ref{sec:bay7be}): the data at thermal neutron energy ($E_{c.m.}$ $=$ $0.0221$~eV) represent four independent measurements of the thermal neutron cross section \citep{koe88,cer89,dam18,tom19}; the red circle and triangle depict the absolute cross section of \citet{mar19} and \citet{gib59}, respectively. For comparison, the light blue curve indicates the energy range important for big bang nucleosynthesis, given by $\sqrt{E_{c.m.}}\exp(-E_{c.m.}/kT)$ at a temperature of $T$ $=$ $0.5$~GK, where $k$ denotes the Boltzmann constant.

The displayed data correspond to the $^7$Be(n,p$_0$)$^7$Li reaction (i.e., the population of the $^7$Li ground state). The only exceptions are the results of \citet{dam18} (red and orange squares) and \citet{koe88} (green inverted triangles). Experimental values of the branching ratio, $\sigma_{np_1}/\sigma_{np_0}$, are in the range of 1.1-2.0\% (see Table~\ref{tab:norm}). Since these values are smaller than the systematic uncertainties of the analyzed data, we will disregard the distinction between the ground-state and the total $^7$Be(n,p)$^7$Li reduced cross section in the statistical analysis. 

Notice that we display in Figure~\ref{fig:data} both statistical and systematic uncertainties for the  absolute reduced cross sections (red data points), and statistical uncertainties only for all other data. The full data set shown includes six absolute cross section normalizations  (with systematic uncertainties provided in parenthesis, see Appendix A): (i) \citet{gib59} (5.0\%); (ii) \citet{mar19} (5.1\%); (iii) \citet{cer89} (8.5\%); (iv) \citet{tom19} (3.2\%); (v) the non-thermal data of \citet{koe88} (2.0\%) (green inverted triangles) are normalized to their thermal cross section (red inverted triangle); (vi) the non-thermal cross sections (orange squares) and the thermal cross section (red square) of \citet{dam18} (10\%) share the same absolute normalization. 

 Figure~\ref{fig:data} raises the fundamental question of how to analyze results from different data sets that are inconsistent with each other within their reported uncertainties. In fact, the $^7$Be(n,p)$^7$Li reaction represents a prime example for discrepant data sets since the reduced cross sections measured by different groups differ by up to $\approx$30\%. We will explain in the next section how these data have been incorporated into our Bayesian model.

\section{Reaction Formalism}\label{sec:rmatrix}
Since we are mainly interested in the low-energy region, with center-of-mass neutron energies below $E_{^7\rm{Be+n}}$ $\approx$ $420$~keV, we will follow previous works \citep{mac58,des04,dam18} and describe the theoretical cross section using an incoherent sum of one-level, two-channel R-matrix approximation expressions. 

The angle-integrated cross section for an isolated resonance is given by \citep{lan58}
\begin{equation}
\sigma_{np}(E) = \frac{\pi}{k^2} \frac{2J+1}{(2j_1 + 1)(2j_2 + 1)} \frac{\Gamma_n \Gamma_p}{(E_0 + \Delta - E)^2 + (\Gamma/2)^2},
\label{eq:eq1}
\end{equation}
where $k$ and $E$ are the wave number and energy, respectively, in the $^7$Be $+$ $n$ center-of-mass system, $J$ is the resonance spin, $j_1$ $=$ $1/2$ and $j_2$ $=$ $3/2$ are the spins of the neutron and $^7$Be, respectively, and $E_0$ is the level eigenenergy. The partial widths of the $^7$Be $+$ $n$ and $^7$Li $+$ $p$ channels ($\Gamma_n$, $\Gamma_p$), the total width ($\Gamma$), and total level shift ($\Delta$), which are all energy dependent, are given by 
\begin{equation}\label{eq:gamwidth}
\Gamma = \sum_c \Gamma_c = \Gamma_n + \Gamma_p~,~~~\Gamma_c = 2 \gamma_c^2 P_c,
\end{equation}
\begin{equation}
\Delta = \sum_c \Delta_c = \Delta_n + \Delta_p~,~~~\Delta_c = - \gamma_c^2 (S_c - B_c),
\label{eq:shift}
\end{equation}
where $\gamma_c^2$ is the reduced width, and $B_c$ is the boundary condition parameter. The energy-dependent quantities $P_c$ and $S_c$ denote the penetration factor and shift factor for channel $c$ (either $^7$Be $+$ $n$ or $^7$Li  $+$ $p$). They are computed numerically from the Coulomb wave functions, $F_\ell$ and $G_\ell$, according to 
\begin{equation}
P_c = \frac{k a_c}{F_\ell^2 + G_\ell^2}~,~~~S_c = \frac{k a_c(F_\ell F_\ell^\prime+G_\ell G_\ell^\prime)}{F_\ell^2 + G_\ell^2}.
\end{equation}
The Coulomb wave functions and their radial derivatives are evaluated at the channel radius, $a_c$, and the quantity $\ell$ denotes the orbital angular momentum for a given channel.  Since we are not using the Thomas approximation \citep{tho51}, all of our partial and reduced widths are `formal' R-matrix parameters \citep[e.g.,][]{des10}. 

By investigating the strength of the residual interaction in nuclei, \citet{dov69} found for the reduced width, $\gamma^2_{\lambda c}$, of an eigenstate, $\lambda$, a limit of
\begin{equation}
\gamma^2_{\lambda c} \lesssim \frac{\hbar^2}{m_c a_c^2},
\label{eq:wl}
\end{equation}
for an individual resonance in a {\it nucleon} channel. The quantity $\gamma^2_{\mathrm{ WL}}$ $\equiv$ $\hbar^2/(m_c a_c^2)$ is often referred to as the Wigner limit \citep{tei52}, where $m_c$ denotes the reduced mass of the interacting particle pair in channel $c$. Considering the various assumptions made in deriving the above expressions, $\gamma^2_{\mathrm{ WL}}$ provides nothing more than an approximation for the upper bound of a reduced width. 

R-matrix parameters and cross sections derived from data have a well-known dependence on the channel (or interaction) radius, which is usually expressed as
\begin{equation}
\label{eq:radius}
a_c = r_0 \left(  A^{1/3}_{1} +  A^{1/3}_{2}  \right),
\end{equation}
where $A_i$ are the mass numbers of the interacting nuclei, and $r_0$ is the radius parameter, with a value usually chosen in the vicinity of $\approx$ $1.4$~fm. The channel radius dependence arises from the truncation of the R-matrix to a restricted number of poles (i.e., a finite set of eigenenergies). The radius of a given channel has no rigorous physical meaning, except that the chosen value should exceed the sum of the radii of the colliding nuclei \citep[e.g.,][and references therein]{des10}. The radius dependence can likely be reduced by including more levels (including background poles) in the data analysis, but only at the cost of increasing the number of fitting parameters. 

The last point to be addressed here is the arbitrary choice of the boundary condition parameter, $B_c$. The eigenenergy, $E_0$, and the reduced widths, $\gamma^2_n$ and $\gamma^2_p$, depend on the values of $B_c$, as is evident from their definitions. Specifically, it can be seen from Equations~(\ref{eq:eq1}) and (\ref{eq:shift}) that changing $B_c$ will result in a corresponding change of $E_0$ to reproduce the measured location of the cross section maximum. Frequently, the maximum occurs when the first term in the denominator of Equation~(\ref{eq:eq1}) is set equal to zero. In that case, a resonance energy, $E_r$, can be defined by 
\begin{equation}\label{eq:narrow}
E_0 + \Delta(E_r) - E_r \equiv 0. 
\end{equation}
One, but not the only, choice for the boundary condition parameter is then $B_c$ $=$ $S_c(E_r)$, which results in $\Delta(E_r)$ $=$ $0$, or $E_r$ $=$ $E_0$. This choice is commonly adopted in the literature, and we will also use it in our analysis. \citet{bar72} showed that the choice of boundary condition parameters does not affect the fitted cross section, i.e., {\it ``any fit to the data with a particular set of $B_c$ values can be duplicated exactly for any other set of $B_c$ values.''} 

We have already introduced in Section~\ref{sec:data} the reduced cross section,  $S_{np}(E_{c.m.})$ $\equiv$ $\sqrt{E_{c.m.}}\sigma_{np}(E_{c.m.})$, which removes the $1/v$ dependence of the reaction cross section for s-wave neutrons. A constant reduced cross section is seen in Figure~\ref{fig:data} below an energy of $10^{-4}$~MeV, indicating that the $^7$Be(n,p)$^7$Li reaction is dominated by s-wave neutrons at those low energies.

\section{Statistical Model Formalism}
\label{sec:bayes}
\subsection{Bayesian inference}
The hierarchical Bayesian model applied in the present work is similar to those presented in \citet{ili16,gom17,des19a,des19b}, to which the reader is referred for more information. However, our model differs in important details from the earlier approaches because of some peculiarities of the $^7$Be(n,p)$^7$Li reaction. We will first give a brief overview of the statistical framework and then present our full model. 

Bayes' theorem is given by \citep{jay03,2017bmad}
\begin{equation}
p(\theta|y) = \frac{\mathcal{L}(y|\theta)\pi(\theta)}{\int \mathcal{L}(y|\theta)\pi(\theta)d\theta},
\label{eq:bayes}
\end{equation}
where the data are denoted by the vector $y$ and the complete set of model parameters is given by the vector $\theta$. All factors entering in Equation~(\ref{eq:bayes}) represent probability densities: $\mathcal{L}(y|\theta)$ is the likelihood, i.e., the probability that the data, $y$, were obtained assuming given values of the model parameters, $\theta$; $\pi(\theta)$ is called the prior, which represents our state of knowledge about each parameter before analyzing the data; the product of likelihood and prior defines the posterior, $p(\theta|y)$, i.e., the probability of obtaining the values of a specific set of model parameters given the data; the denominator, called the evidence, is a normalization factor and is not important in the context of the present work. It is apparent from Equation~(\ref{eq:bayes}) that the posterior represents an update of our prior state of knowledge about the model parameters once new data become available.

In the simplest case, when the experimental reduced cross section, $S$ $\equiv$ $\sqrt{E}\sigma$ is subject to statistical uncertainties only, the likelihood is given by
\begin{equation}
\mathcal{L}(S^{\rm exp}|\theta) = \prod_{i=1}^N\frac{1}{\sigma_{\rm stat,i}\sqrt{2\pi}}e^{-\frac{\left[ S^{\rm exp}_i - S(\theta)_i\right]^2}{2\sigma_{\rm stat,i}^2}},
\label{eq:like2}
\end{equation}
where $S(\theta)_i$ is the theoretical reduced cross section (e.g., obtained from R-matrix theory); the product runs over all data points, labeled by $i$. The likelihood represents a product of normal distributions, each with a mean of $S(\theta)_i$ and a standard deviation of $\sigma_{\rm stat,i}$, given by the experimental statistical uncertainty of datum $i$. In symbolic notation, the above expression can be abbreviated by
\begin{equation}
S^{\rm exp}_i \sim \rm{Normal}(S(\theta)_i, \sigma_{stat,i}^2), 
\end{equation}
where ``Normal()'' denotes a normal probability density and the symbol ``$\sim$'' stands for ``has the probability distribution of.'' 

In many cases, the scatter of the measured data cannot be explained solely based on the reported statistical uncertainties, suggesting that additional sources of statistical uncertainties were unknown to the experimenter.  For example, the reported statistical uncertainties may have been too optimistic because target thickness or beam straggling effects were underestimated. We will use the expression {\it extrinsic uncertainty} for describing such effects \citep{des19b}. Since the observed scatter in the data points of a given set contains the information about additional (unknown) statistical effects, our model can predict the magnitude of the extrinsic uncertainty for each data set. If both statistical and extrinsic uncertainties are present in a measurement, the overall likelihood is given by a nested (hierarchical) expression. Using the symbolic notation, we can write  
\begin{align}
&S^\prime_i \sim \mathrm{Normal}(S(\theta)_i, \sigma_{extr}^2),\label{eq:staterror}\\
&S^{\rm exp}_i \sim \mathrm{Normal}(S^\prime_i, \sigma_{stat;i}^2)\label{eq:extrerror}.
\end{align}

These two expressions describe the construction of the overall likelihood and have the following meaning: first, an unknown source of statistical scatter, quantified by the standard deviation $\sigma_{extr}$ of a normal probability density, perturb the true (but unknown) value of the reduced cross section at energy $i$, $S(\theta)_i$, to produce a value of $S_i^\prime$; second, the latter value is perturbed, in turn, by the experimental statistical uncertainty, quantified by the standard deviation $\sigma_{\rm stat,i}$ of a normal probability density, to produce the measured value of $S^{exp}_i$. The above example demonstrates how experimental effects impacting the data can be implemented in a straightforward manner into a Bayesian model. 

Each of the model parameters contained in the vector $\theta$ requires a prior. It contains the information on the probability density of a given parameter prior to analyzing the data under consideration. Priors must be chosen to best represent the physics involved. For example, if a $^8$Be level corresponding to a $^7$Be $+$ $n$ resonance energy of $E_r^{exp} \pm \Delta E_r^{exp}$ has been observed above the neutron threshold, in a reaction other than $^7$Be(n,p)$^7$Li, we can write for the prior of the resonance energy
\begin{equation}
E_r \sim \rm{TruncNormal}(E_r^{exp}, \left[ \Delta E_r^{exp}\right]^2), 
\end{equation}
where we assume that the standard deviation of the normal density is equal to the resonance energy uncertainty and the normal density is truncated at zero energy to exclude negative values.

Systematic uncertainties require special treatment. If an experimenter reports a systematic uncertainty, for example, of $\pm5\%$, we may assume that the systematic factor uncertainty is $1.05$. The true value of the normalization factor, $f$, is unknown, otherwise we would have corrected for the effect and there would be no systematic uncertainty. This implies that we do have one piece of information: the expectation value of the normalization factor is unity. A useful distribution for normalization factors is the lognormal probability density, which is characterized by two quantities, the location parameter, $\mu$, and the shape parameter, $\sigma$. The median value of the lognormal distribution is given by $x_{med}$ $=$ $e^\mu$, while the factor uncertainty, for a coverage probability of 68\%, is $f.u.$ $=$ $e^\sigma$. We will include in our Bayesian model a systematic effect on the reduced cross section as an informative, lognormal prior with a median of $x_{med}$ $=$ $1.0$ (or $\mu$ $=$ $\ln x_{med}$ $=$ $0$), and a factor uncertainty given by the systematic uncertainty, i.e., in the above example, $f.u.$ $=$ $1.05$ (or $\sigma$ $=$ $\ln f.u.$ $=$ $\ln (1.05)$). The prior is then explicitly given by
\begin{equation}\label{eq:logn1}
\pi(f) = \frac{1}{ \ln (f.u.) \sqrt{2 \pi} f} 
e^{-\frac{[\ln f]^2}{2[\ln(f.u.)]^2}}, 
\end{equation}
or 
\begin{equation}\label{eq:logn2}
f \sim \rm{LogNormal}(0, [\ln(f.u.)]^2).
\end{equation}
where ``LogNormal'' denotes a lognormal probability density. For more information on this choice of prior, see \citet{ili16}. 

In conventional $\chi^2$-square fitting, normalization factors are viewed as a systematic shift in the {\it data}. In the Bayesian model, the reported data are not modified. Instead, during the fitting, each data set ``pulls'' on the true reduced cross section curve with a strength inversely proportional to the systematic uncertainty: a data set with a small systematic uncertainty will pull the true S-factor curve more strongly towards it than a data set with a large systematic uncertainty.

\subsection{Bayesian model for \textsuperscript{7}B\MakeLowercase{e(n,p)}\textsuperscript{7}L\MakeLowercase{i}}
\label{sec:bay7be}
Our model includes the following parameters: (i) R-matrix parameters, i.e., the energy ($E_r$) of each resonance, reduced neutron and proton widths ($\gamma^2_d$, $\gamma^2_n$) for each level, and the neutron and proton channel radii ($a_n$, $a_p$); (ii) for each data set, the extrinsic scatter for the reduced cross section ($\sigma_{extr}$), and the reduced cross section normalization factors ($f$, $g$; see below). As discussed in Section~\ref{sec:data}, we consider results from six experiments, providing 160 data points. Statistical uncertainties are assumed to be normally distributed. Experimental mean values for the measured energies and reduced cross sections, together with estimates of statistical and systematic uncertainties, are given in Appendix \ref{sec:app}. We will discuss below the priors for the physical model parameters (resonance energies, reduced widths, and channel radii) and the experimental model parameters (data uncertainties and normalization factors).

The adopted priors for the resonance energies are discussed in Section~\ref{sec:structure} and are listed in the last column of Table~\ref{tab:nuclear}. Normal densities truncated at zero are assumed for the reduced widths ($\gamma^2_d$ and $\gamma^2_n$), with standard deviations given by 50\% of the Wigner limits ($\gamma^2_{\rm WL,n}$ and $\gamma^2_{\rm WL,p}$) for the neutron and proton channel. This choice of prior takes into account the approximate nature of the Wigner limit concept (Equation~(\ref{eq:wl})). 

\citet{des10} recommended to choose the channel radius so that its value exceeds the sum of the radii of the colliding nuclei. Most previous studies of the $^7$Be $+$ $n$ and $^7$Li $+$ $p$ reactions adopted {\it ad hoc} values. \citet{mac58} assumed a radius parameter of $1.45$~fm, which results in a channel radius of $4.2$~fm, according to Equation~(\ref{eq:radius}). In the R-matrix study of \citet{bar77}, values of $a_n$ $=$ $a_p$ $=$ $4.2$~fm and $a_n$ $=$ $3.86\pm0.15$~fm are reported, but no information is provided about how the latter value and its uncertainty were determined. The multi-channel R-matrix study of \citet{koe88} used values of $a_n$ $=$ $a_p$ $=$ $3.0$~fm, stating {\it ``...the automated fitting procedure at the value (3 fm) of the channel radii that it preferred.''} However, it is not clear if the channel radii were fit parameters or if they were kept constant. The $^7$Be(n,p)$^7$Li R-matrix study of \citet{ada03} does not mention any value for the channel radius, although a value of $5$~fm is stated in \citet{des04}. The comprehensive R-matrix fit of \citet{pag05} employed fixed $^7$Be $+$ $n$ and $^7$Li $+$ $p$ channel radii of $3.0$~fm, which are {\it ``based on earlier R-matrix analyses.''} In the present work, we will choose for the channel radii normal priors with a mean of $4.0$~fm and a standard deviation of $0.5$~fm, which are truncated at zero to exclude negative values.

Before we choose priors for the experimental model parameters, the following problem needs to be addressed. How can we combine data sets of very different sizes in a comprehensive fit? For instance, Figure~\ref{fig:data} shows many data points from the experiment of \citet{koe88} and a single datum from the measurement of \citet{cer89}. The standard approach of fitting both data sets together, irregardless of their size, will render the single datum of the latter work irrelevant. 

Recall that we adopted the results from six independent experiments. To treat all of these measurements in a consistent manner, we started from the assumption that each independent measurement generally provides two pieces of information: (i) the reduced cross section normalization (i.e., the absolute cross section), and (ii) the energy dependence of the reduced cross section (i.e., the relative cross section). 

For example, as already pointed out, \citet{koe88} normalized their relative cross section at non-thermal energies to their measured absolute cross section at thermal neutron energy. Therefore, we adopt the informative prior of Equations~(\ref{eq:logn1}) and (\ref{eq:logn2}) for the datum at thermal neutron energy. Since \citet{koe88} report a systematic uncertainty of 2.0\% (see Table~\ref{tab:norm} and Appendix~\ref{sec:koehlerdata}), we assume for the prior of their normalization factor 
\begin{equation}
f_{\rm Koehler} \sim \rm{LogNormal}(0, [\ln(1.020)]^2).
\end{equation}
The thermal cross section also has a statistical uncertainty (see Table~\ref{tab:norm}), which is included in the likelihood according to Equation~(\ref{eq:staterror}). Since the thermal cross section was measured, by definition, at a single energy only, it has no extrinsic (i.e., additional) scatter. Also, the single datum cannot provide information on the energy dependence of the reduced cross section.

On the other hand, for the non-thermal (i.e., relative) cross sections of \citet{koe88} we chose to scale the true (unknown) cross section  by a factor of $10^g$, with a non-informative prior of 
\begin{equation}
g_{\rm Koehler} \sim \rm{Uniform}(-1,1), 
\end{equation}
corresponding to a uniform prior between $-1$ and $1$. In this case, the normalization factor, $10^g$, is varied by up to one order of magnitude up or down during the sampling. Therefore, the (relative) non-thermal energy data points provide only information on the energy dependence of the reduced cross section, but no information on the absolute normalization. The non-thermal data points also have individual statistical uncertainties and a common extrinsic uncertainty, which are included in the likelihood according to Equations~(\ref{eq:staterror}) and (\ref{eq:extrerror}). 

We proceeded in a similar fashion with all the other data sets. The single data points of \citet{cer89} and \citet{tom19} at thermal neutron energy provide only information on the absolute cross section. The data sets of \citet{gib59} and \citet{mar19} were also split into two parts, one part containing all data points, except one, providing only relative cross section information, and one part containing a single data point (here arbitrarily chosen at the lowest measured energy) that provides only information on the absolute normalization. Since the thermal and non-thermal data of \citet{dam18} share the same absolute normalization, we treated their reported thermal cross section as an absolute cross section, and all other data points as relative cross sections.

The extrinsic uncertainties of the measured cross sections are inherently unknown to the experimenter. Thus we will adopt in this case broad normal priors that are truncated at zero, with standard deviations of $2$~$\sqrt{\mathrm{MeV}}$b. 

Our complete Bayesian model is summarized below in symbolic notation:
\begin{alignteo}
\label{eq:model}
    & \textrm{{\bf Model relationship:}} \notag \\
    & \indent S_i(E) = B + \sqrt{E} \sum_{r} \sigma_{np;r}(E).\notag \\     
    & \textrm{{\bf Parameters:}} \notag \\
     & \indent \theta \equiv (E_r,\gamma^2_{n;r},\gamma^2_{p;r}, a_n, a_p, \sigma_{extr,j},g_{j},f_{k}). \notag \\    
%
    & \textrm{{\bf Likelihood (relative data):}} \notag \\ 
    & \indent S^{\prime}_{i,j} = 10^{g_{j}} \times S_i, \notag \\
    & \indent S^{\prime\prime}_{i,j}  \sim \rm{Normal}(S^{\prime}_{i,j}, \sigma_{extr,j}^2), \notag \\
    & \indent S^{exp}_{i,j}  \sim \rm{Normal}(S^{\prime\prime}_{i,j}, \sigma_{stat,i}^2). \notag \\
    & \textrm{{\bf Likelihood (absolute data):}}\\ 
    & \indent S^{\prime}_{i,k} = f_{k} \times S_i, \notag \\
    & \indent S^{exp}_{i,k}  \sim \rm{Normal}(S^{\prime}_{i,k}, \sigma_{stat,i}^2). \notag \\
    & \textrm{{\bf Priors:}}\notag \\ 
    & \indent E_r \sim \rm{TruncNormal}(E_r^{exp}, [\Delta E_r^{exp}]^2),	\notag \\
    & \indent (\gamma_{n;r}^2, \gamma_{p;r}^2) \sim  \rm{TruncNormal}(0, [0.5\gamma_{WL}^2]^2),   \notag \\
    & \indent (a_n, a_p) \sim \rm{TruncNormal}(4.0, 0.5^2), \notag \\
    & \indent \sigma_{\mathrm{extr},j} \sim \rm{TruncNormal}(0, 2.0^2), \notag \\ 
    & \indent f_{k} \sim \mathrm{LogNormal}(0, [\ln(f.u.)_k]^2),  \notag \\
    & \indent g_{j} \sim \rm{Uniform}(-1,1),  \notag \\
    & \indent B \sim \rm{TruncNormal}(0, 0.5^2). \notag 
\end{alignteo}
The index $i = 1,...,160$ labels individual data points, $j = 1,...,4$ denotes the relative data sets with information on the energy dependence only \citep{gib59,koe88,mar19,dam18}, $k$ $=$ $1,...,6$ labels the absolute data sets with information on the cross section normalization only \citep{koe88,dam18,gib59,mar19,cer89,tom19}, and $r$ $=$ $1,...,7$ denotes the resonances.
The symbols have the following meaning: measured energy ($E_r^{exp}$) and measured reduced cross section ($S^{exp}$); true resonance energy ($E_r$); the true reduced cross section ($S$) is calculated from the cross section, $\sigma_{np}$, according to Equation~(\ref{eq:eq1}), using the R-matrix parameters ($E_r$, $\gamma^2_{n;r}$, $\gamma^2_{p;r}$, $a_n$, $a_p$). The ``TruncNormal()'' prior refers to a truncated normal probability distribution, i.e., a density that excludes negative values. The numerical values of the energies ($E$) and reduced widths ($\gamma^2_n$, $\gamma^2_p$) are in units of MeV,  the channel radii ($a_n$, $a_p$) are in units of fm, and the values of the reduced cross sections ($S$), extrinsic scatters ($\sigma_{extr}$), and statistical uncertainties ($\sigma_{stat}$) are in units of $\sqrt{\mathrm{MeV}}$b. The quantity $B$ is a constant added to the total reduced cross section to account for contributions from higher-lying resonances.

\section{Nuclear structure of $^8$B\MakeLowercase{e} near the neutron threshold}
\label{sec:structure}
In our analysis, we are taking into account the seven lowest-lying $^8$Be levels above the neutron threshold most relevant for the $^7$Be(n,p)$^7$Li reaction rate. They are listed in Table~\ref{tab:nuclear}. The excitation energies given in column 1, which are most frequently quoted in the literature, are adopted from ENSDF. 
Four of the seven levels have no assigned excitation energy uncertainty. We already pointed out in Section~\ref{sec:intro} that in some previous fits \citep{kun16,dam18} the resonance energies were kept fixed.
We suspect that, if nuclear data evaluators do not provide an uncertainty, it can be reasonably assumed that the reported mean values have significant uncertainties. All of these levels have a significant total width, which partly explains the difficulty in assigning uncertainties to the excitation energies. We will evaluate in the following the nuclear structure information previously reported for the levels listed in Table~\ref{tab:nuclear}. Our goal is to estimate reasonable probability densities for the priors of the resonance energies, which are listed in the last column of Table~\ref{tab:nuclear}.

\subsection{The 2$^-$ level near $E_x$ $\approx$ $18.9$~MeV}\label{sec:2-}
A 2$^-$ level in $^8$Be near the neutron threshold, corresponding to a s-wave resonance in $^7$Be(n,p)$^7$Li, was first suggested by \citet{bre48}. The $^7$Li(p,$\gamma$)$^8$Be$^\ast$ $\rightarrow2\alpha$ measurement by \citet{swe69} located the level at E$_x$ $=$ $18.9$~MeV excitation energy, just above the neutron threshold ($S_n$ $=$ 18898.64$\pm$0.08~keV; see Table~\ref{tab:nuclear}), with a suggested total width of $\Gamma$ $=$ $150$ $\pm$ $50$~keV. The R-matrix analysis of \citet{koe88} predicted a value of E$_x$ $=$ $18.89$~MeV, just below threshold, and a width of $\Gamma$ $=$ $122$~keV. The R-matrix analysis of \citet{ada03} found a value of E$_x$ $=$ $18.901$~MeV. They kept the partial widths, which were adopted from the literature, fixed in their fit. Finally, a comprehensive multi-channel R-matrix study including 69 experimental references reported values of E$_x$ $=$ $18.92$~MeV and $\Gamma$ $=$ $120$~keV \citep{pag05}. The value of ``E$_x$ $=$ $18910$~keV'' listed in ENSDF was first mentioned in Table~8.13 of  \citet{lau66} and was carried through without updating it using newer experimental information.

The case is exacerbated by the fact that the energies and widths reported in the above studies have different meanings, depending on the details of the applied nuclear reaction model (e.g., single or multi-level R-matrix analysis, Breit-Wigner expressions, etc.). We will account for the significant uncertainty in the excitation energy and total width of the 2$^-$ level by locating it at the neutron threshold and by assuming a broad prior, $E_r \sim \rm{TruncNormal}(0, 0.10^2)$, i.e., a truncated normal distribution with a peak at $E_r$ $=$ $0$~keV and a standard deviation of $100$~keV.

\subsection{The 3$^+$ level near $E_x$ $\approx$ $19.1$~MeV}
\citet{rie63} observed a resonance in the $^7$Li(p,$\gamma_1$)$^8$Be reaction at E$_R^{lab}$ $=$ $2.06\pm0.02$~MeV, which corresponds to an excitation energy of E$_x$ $=$ $19056\pm17$~keV, and determined a total width of $\Gamma$ $=$ $271\pm18$~keV. By measuring the $^9$Be(d,t)$^8$Be reaction, \citet{oot77} reported values of E$_x$ $=$ $19071\pm10$~keV and $\Gamma$ $=$ $270\pm30$~keV. The weighted average of these results is listed in ENSDF. However, while a clear peak is observed in the yield curve measured by \citet{rie63}, there is no convincing evidence of a peak in the $^9$Be(d,t)$^8$Be spectra presented in \citet{oot77} (see their Figures~2b and 7). Furthermore, the comprehensive multi-level R-matrix analysis of \citet{pag05} finds E$_x$ $=$ $19.02$~MeV and $\Gamma$ $=$ $270$~keV. Therefore, it appears that the excitation energy uncertainty of $10$~keV reported by \citet{oot77} is too optimistic.

For the values quoted above, we find an average excitation energy of $19050$~keV, corresponding to a $^7$Be(n,p)$^7$Li center-of-mass resonance energy of $150$~keV. Based on the available information, we estimate a value of $25$~keV for the uncertainty in the excitation energy.  Hence, we will adopt for the prior $E_r$ $\sim$ $\rm{TruncNormal}(0.150, 0.025^2)$.

\subsection{The 3$^+$ level near $E_x$ $\approx$ $19.2$~MeV}
Measured peaks in pulse-height spectra caused by the decay of this level were clearly observed in a number of works. \citet{ajz76} reported a value of E$_x$ $=$ $19220\pm30$~keV by measuring the $^9$Be($^3$He,$\alpha$)$^8$Be reaction. \citet{oot77} found values of E$_x$ $=$ $19261\pm32$~keV and $\Gamma$ $=$ $220\pm30$~keV from a $^9$Be(d,t)$^8$Be reaction study. \citet{hei89} measured the Ag($^{14}$N,$^8$Be) reaction and reported values of E$_x$ $=$ $19234\pm12$~keV and $\Gamma$ $=$ $210\pm35$~keV. These values were used to derive the weighted averages, E$_x$ $=$ $19235\pm10$~keV and $\Gamma$ $=$ $227\pm16$~keV, listed in ENSDF. This excitation energy corresponds to a resonance energy of $336$~keV. For the prior, we will assume  $E_r \sim \rm{TruncNormal}(0.336, 0.010^2)$.

\subsection{The 1$^-$ level near $E_x$ $\approx$ $19.4$~MeV}
The situation regarding this $^8$Be level is ambiguous. A 1$^-$ resonance near a laboratory energy of $2.5$~MeV in the $^7$Li(p,n$^\prime$)$^7$Be$^\ast$ reaction, corresponding to a $^8$Be level at $19.4$~MeV, was reported by \citet{bev61}, but was later questioned in \citet{buc64}. A ``knee'' in the $^7$Li(p,$n_1$)$^7$Be excitation function led \citet{ple72} to suggest an s-wave (1$^-$) resonance near a center-of-mass proton energy of $2.30$~MeV, resulting in an $^8$Be excitation energy of $E_x$ $\approx$ $19.5$~MeV and a width of $\Gamma$ $\approx$ $0.65$~MeV. \citet{bro73} measured the $^7$Li($\overrightarrow{p}$,p)$^7$Li reaction and their phase shift analysis found hints of a 1$^-$ resonance at a laboratory resonance energy near $2.5$~MeV. The multi-level R-matrix study of \citet{pag05} locates a 1$^-$ level at $E_x$ $=$ $19.33$~MeV with a width of $\Gamma$ $=$ $650$~keV. 

Although the existence of a 1$^-$ level near $19.4$~MeV excitation energy is questionable at present, we adopt tentatively an average excitation energy of E$_x$ $\approx$ $19410\pm100$~keV, corresponding to a center-of-mass resonance energy of $510$~keV. We will assume  $E_r \sim \rm{TruncNormal}(0.51, 0.10^2)$ for the prior.

\subsection{The 4$^+$ level near $E_x$ $\approx$ $19.9$~MeV}\label{sec:4+}
\citet{bac72} studied $\alpha\alpha$ scattering and observed a rapid rise in the $\ell$ $=$ $4$ phase shift, corresponding to a 4$^+$ level near $19.8$~MeV excitation in $^8$Be. The total width was estimated as {\it ``less than 1 MeV.''} \citet{oot77} observed a very weak and broad peak at $E_x$ $=$ $19.86\pm0.05$~MeV, with a width of $\Gamma$ $=$ $700\pm100$~keV. Considering the small magnitude of the peak shown in their $^9$Be(d,t)$^8$Be pulse-height spectrum (see their Figure~7), we find their reported uncertainty too optimistic. We will adopt their excitation energy, corresponding to a center-of-mass resonance energy of $960$~keV, with an uncertainty of $100$~keV. Therefore, we assume  $E_r \sim \rm{TruncNormal}(0.96, 0.10^2)$ for the prior. 

\subsection{The $E_x$ $\approx$ $20.1$~MeV (2$+$) and $E_x$ $\approx$ $20.2$~MeV (0$^+$) levels}
The $\alpha\alpha$ elastic scattering study of \citet{bac72} reported a 2$^+$ level at $20.2$~MeV, and a 0$^+$ level near $20.3$~MeV with a level width of less than $1$~MeV. The location of the former level is consistent with an R-matrix analysis of $^7$Li(p,$\alpha$)$^4$He data \citep{kum71}, which reported a level energy of $20.1$~MeV. The $^7$Li(d,$\alpha\alpha$)n study of \citet{are91} obtained total widths of $\Gamma$ $=$ (0.85$\pm$0.25)~MeV (2$^+$) and $\Gamma$ $=$ (0.75$\pm$0.25)~MeV (0$^+$). Notice that the small uncertainties of the total widths of these two levels listed in ENSDF are erroneous. The R-matrix analysis of \cite{pag05} resulted in values of $E_x$ $=$ $20.10$~MeV, $\Gamma$ $=$ $680$~keV (2$^+$) and $E_x$ $=$ $20.13$~MeV, $\Gamma$ $=$ $750$~keV (0$^+$).

The averages of the excitation energies result in $^7$Be(n,p)$^7$Li center-of-mass resonance energies of $1230$~keV (2$^+$) and  $1320$~keV (0$^+$). No reliable estimates exist for the excitation energy uncertainties. We will adopt in the present work priors of $E_r \sim \rm{TruncNormal}(1.23, 0.10^2)$ (2$^+$) and $E_r \sim \rm{TruncNormal}(1.32, 0.10^2)$ (0$^+$).

\begin{deluxetable*}{lccccl}
\tablecaption{Information on Astrophysically Relevant $^8$Be Levels Above the Neutron Threshold Prior to the Present Work.}
\label{tab:nuclear} 
\tablewidth{\columnwidth}
\tablehead{
$E_x$ (keV)\tablenotemark{a}   		&     $E_x$ (keV)\tablenotemark{b}    	&     $J^\pi$ \tablenotemark{a}    	&   $\ell$\tablenotemark{c}	&    $\Gamma$ (keV) \tablenotemark{b}     &   Prior \tablenotemark{d}   
} 
\startdata
18910	& 18900, 18890, 18920 & 2$^-$& 0& 150$\pm$50, 122, 120&  $\rm{TruncNormal}(0,0.10^2)$ \\   
19069$\pm$10& 19056$\pm$17, 19071$\pm$10, 19020 & 3$^+$	& 1	& 271$\pm$18, 270$\pm$30, 270 &       $\rm{TruncNormal}(0.150, 0.025^2)$ \\   
19235$\pm$10 & 19220$\pm$30, 19261$\pm$32, 19234$\pm$12 & 3$^+$	& 1	& 220$\pm$30, 210$\pm$35&       $\rm{TruncNormal}(0.336, 0.010^2)$     \\   
19400& (19400), (19500), (19330) & 1$^-$& 0	& (650)&$\rm{TruncNormal}(0.51, 0.10^2)$\\   
19860$\pm$50&  19860$\pm$50, 19800& 4$^+$ & 3&  $<$1000, 700$\pm$100&  $\rm{TruncNormal}(0.96, 0.10^2)$       \\   
20100 & 20200, 20100, 20100	&  2$^+$ & 1 &  850$\pm$250, 680 & $\rm{TruncNormal}(1.23, 0.10^2)$ \\   
20200 & 20300, 20130 & 0$^+$ & 1 & 750$\pm$250, $<$1000, 750 & $\rm{TruncNormal}(1.32, 0.10^2)$  \\   
\enddata
\tablenotetext{a}{From ENSDF.}
\tablenotetext{b}{Excitation energies and total widths reported in the original literature (see text).}
\tablenotetext{c}{Orbital angular momentum in the entrance and exit channel; only the lowest value is listed if more than one value is allowed.}
\tablenotetext{d}{Prior for resonance energy. The  first value indicates the mode of a normal density truncated at zero energy ($\rm{TruncNormal}$), corresponding to our best estimate of the resonance energy, $E_r$ (in MeV); the second value represents the variance, corresponding to our best estimate of the resonance energy uncertainty, $\Delta E_r$ (in MeV). To calculate the resonance energy from the excitation energy we used $S_n$ $=$ 18898.64$\pm$0.08~keV \citep{wan17}. }
\end{deluxetable*}
%
\section{Results}
\label{sec:analysis}
We evaluate the Bayesian model of Equation~(\ref{eq:model}) using an automated factor slice sampler \citep[AFSS;][]{afss}. The AFSS is particularly well suited for our problem since it accounts for potential correlations between the R-matrix parameters. It performs the sampling within a rotated, or ``factor”,  projected space, which yields nearly independent draws even in a scenario of highly correlated parameters and a high-dimensional target distribution. This method was implemented using the {\sc nimble} package \citep{Val2017} within the R language \citep{rcore15}. We randomly sampled all variables of interest using a Markov chain of length 500,000, which included a burn-in phase of 400,000 steps. This ensured convergence of all chains according to the Gelman-Rubin convergence diagnostic \citep{gelman1992}. 

Before comparing our fitted cross sections to previous results, it must be emphasized that our analysis benefits from the availability of a larger body of data, especially regarding absolute thermal cross sections (Table~\ref{tab:norm}). In addition, some of our model assumptions differ from previous works. For example, \citet{koe88} employ different definitions for both the resonance energy and the partial widths, which they define in terms of the properties of an S-matrix pole on a Riemann sheet. Their approach yields a total width much smaller than the sum of the neutron and proton width. In our parameterization, the width of a level is given by Equation~(\ref{eq:gamwidth}). Furthermore, \citet{des04} took only four levels above the neutron threshold into account and kept half of their R-matrix parameters constant in the fitting. \citet{dam18} adopt the level energies from ENSDF and kept all of them constant in their fit. They also did not report uncertainties for their fitted partial widths. Furthermore, unlike our adopted procedure, all previous authors, with one possible exception (Section~\ref{sec:bay7be}), have kept the channel radii at fixed values during the fitting.

\subsection{Reduced cross section}\label{sec:redcross}
\begin{figure*}[hbt!]
\includegraphics[width=1\linewidth]{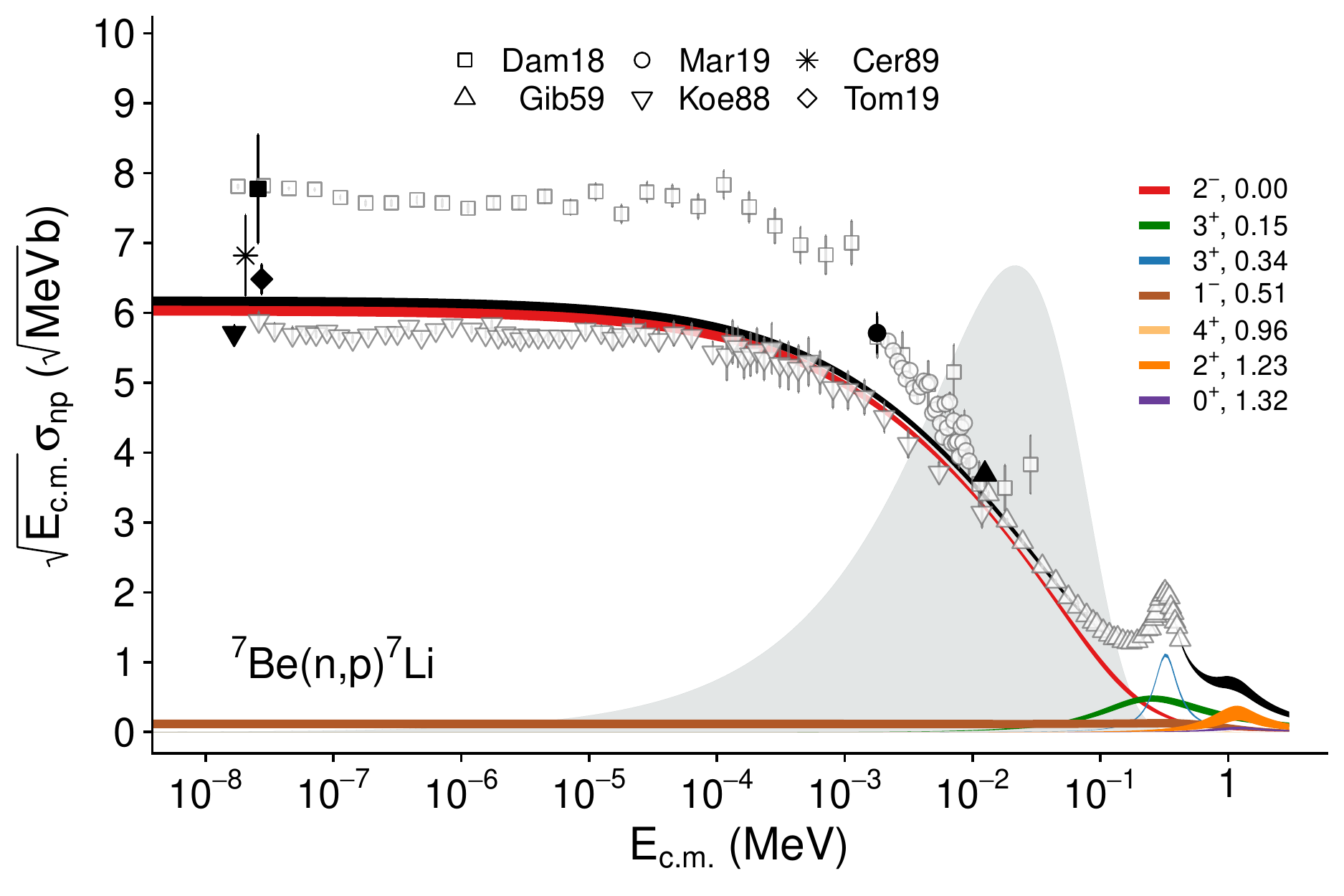}
\caption{\label{fig:MCMCBe7} 
Bayesian R-matrix fit of the reduced total cross section for $^7$Be(n,p)$^7$Li versus the center-of-mass neutron energy (black band). The contributions from the first seven resonances are shown as colored bands. The widths of the bands indicate 68\% credible intervals around the median of the posterior density. The data shown are the same as those displayed in Figure~\ref{fig:data}, with the open gray and full black symbols corresponding to relative and absolute data, respectively.  Data set normalization factors, $f_k$, are not included in any way in the displayed data. The shaded region indicates the energy range important for big bang nucleosynthesis at a temperature of $0.5$~GK. When comparing our fit to previous results, notice that we show the reduced cross section in the center-of-mass system, $\sqrt{E_{c.m.}}\sigma_{np}$, while other authors \citep[e.g.,][]{koe88,dam18} preferred to display the quantity $\sqrt{E_{lab}}\sigma_{np}$.
}
\end{figure*}

Our reduced cross section fit for all data shown in Figure~\ref{fig:data} is displayed in Figure~\ref{fig:MCMCBe7}. The black and colored bands depict the total S-factor and the contributions from individual resonances, respectively. The widths of all bands signify 68\% credible intervals. It can be seen that the 2$^-$ level (red) dominates the total reduced cross section over most of the energy range shown, with a small contribution from the 1$^-$ level (brown). At higher energies, above $E_{c.m.}$ $=$ $0.1$~MeV, the two 3$^+$ states (green and blue) and the 2$^+$ level (orange) contribute to the total reduced cross section. Our fractional resonance contributions agree with the results of \citet{ada03}, who also find that the 2$^-$ level dominates the low-energy cross section. Our results disagree with \citet{dam18}, who reported a $\approx$35\% cross section contribution of the 1$^-$ level at low energies (see their Figure~2). 

The absolute normalization and the energy dependence of the {\it total} reduced cross section is of main interest for the $^7$Be(n,p)$^7$Li reaction rate. At energies between 10$^{-2}$~MeV and 1~MeV, where the fit is determined by the data of \citet{gib59}, our reduced cross section (black band in Figure~\ref{fig:MCMCBe7}) agrees with most previous results. However, at lower energies, our best-fit cross section exceeds the results of \citet{koe88} and \citet{ada03}, and is smaller compared to the result of \citet{dam18}. It can be seen in Figure~\ref{fig:MCMCBe7} that our best fit is pulled more strongly towards the absolute thermal cross sections of \citet{koe88} and \citet{tom19}, with reported systematic uncertainties of 2.0\% and 3.2\% (Table~\ref{tab:norm}), respectively, than to the data of \citet{dam18} (10\%), \citet{cer89} (8.5\%), and \citet{mar19} (5.1\%).

\begin{figure}
\includegraphics[width=1\linewidth]{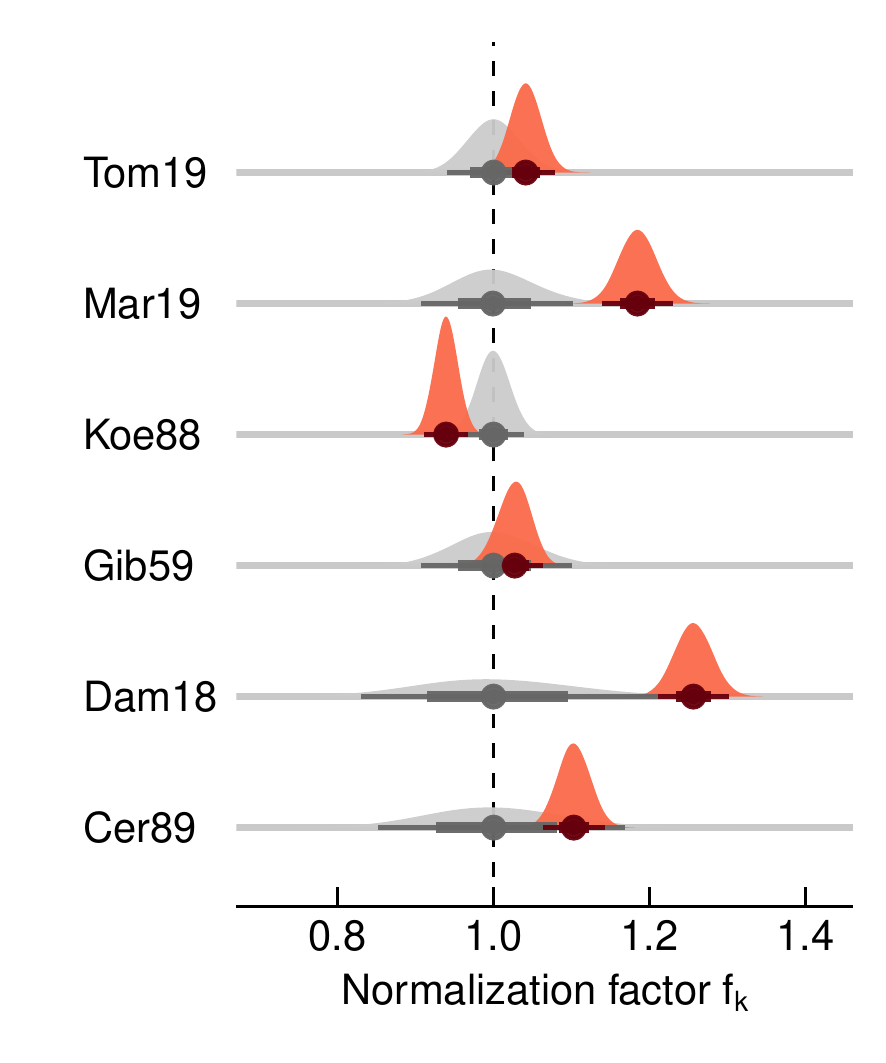}
\caption{Marginalized posteriors (red) of the normalization factors, $f_k$, for each of the six experiments that reported absolute cross sections. The horizontal bar below each distribution shows the median, and the 68\% and 95\% credible intervals. For comparison, the corresponding prior distributions are displayed in gray. For the data set labels, see Figure \ref{fig:MCMCBe7}.}
\label{fig:Be7_norm}
\end{figure}

The posteriors of the normalization factors, $f_k$, for each of the experiments that reported absolute cross sections are displayed in Figure~\ref{fig:Be7_norm} as red areas. For comparison, the densities shown in gray show the priors,  with their spreads determined by the reported systematic uncertainties, according to Equation~(\ref{eq:model}). Our numerical values are $f_{\rm Koe88}$ $=$ 0.94$\pm$0.01, $f_{\rm Dam18}$ $=$ 1.26$\pm$0.02, $f_{\rm Gib59}$ $=$ 1.03$\pm$0.02, $f_{\rm Mar19}$ $=$ 1.18$\pm$0.02, $f_{\rm Cer89}$ $=$ 1.10$\pm$0.02, and $f_{\rm Tom19}$ $=$ 1.04$\pm$0.02. Recall that these values represent factors by which the true cross section is multiplied to agree with the data, as explained in Section~\ref{sec:bayes}. 

The largest deviations from a normalization of unity are exhibited by the data sets of \citet{dam18} and \citet{mar19}. In both cases, the sampled normalization factors are much larger than their reported systematic uncertainties. It appears that these authors may have underestimated the impact of systematic effects on their experiment.

 For the spread parameter of the extrinsic scatter, $\sigma_{\rm extr}$, we find the following values: $0.105\pm0.028$ MeV b \citep{dam18},  $0.0123\pm0.0052$ MeV b \citep{gib59},  
$0.124\pm0.005$ MeV b \citep{mar19}, and 
$0.0192\pm0.0138$ MeV b \citep{koe88}, respectively. The largest scatter is present in the data of \citet{mar19} and \citet{gib59}. 
For comparison, the mean and standard deviation of the reported statistical uncertainties are: $0.168\pm0.126$ MeV b \citep{dam18},  $0.0176\pm0.0049$ MeV b \citep{gib59},  
$0.120\pm0.034$ MeV b \citep{mar19}, and 
$0.159\pm0.109$ MeV b \citep{koe88}. Hence, the extrinsic scatter is either smaller or comparable in magnitude to the reported statistical uncertainties.

\subsection{R-matrix parameters}
\begin{figure*}[hbt!]
\includegraphics[width=1\linewidth]{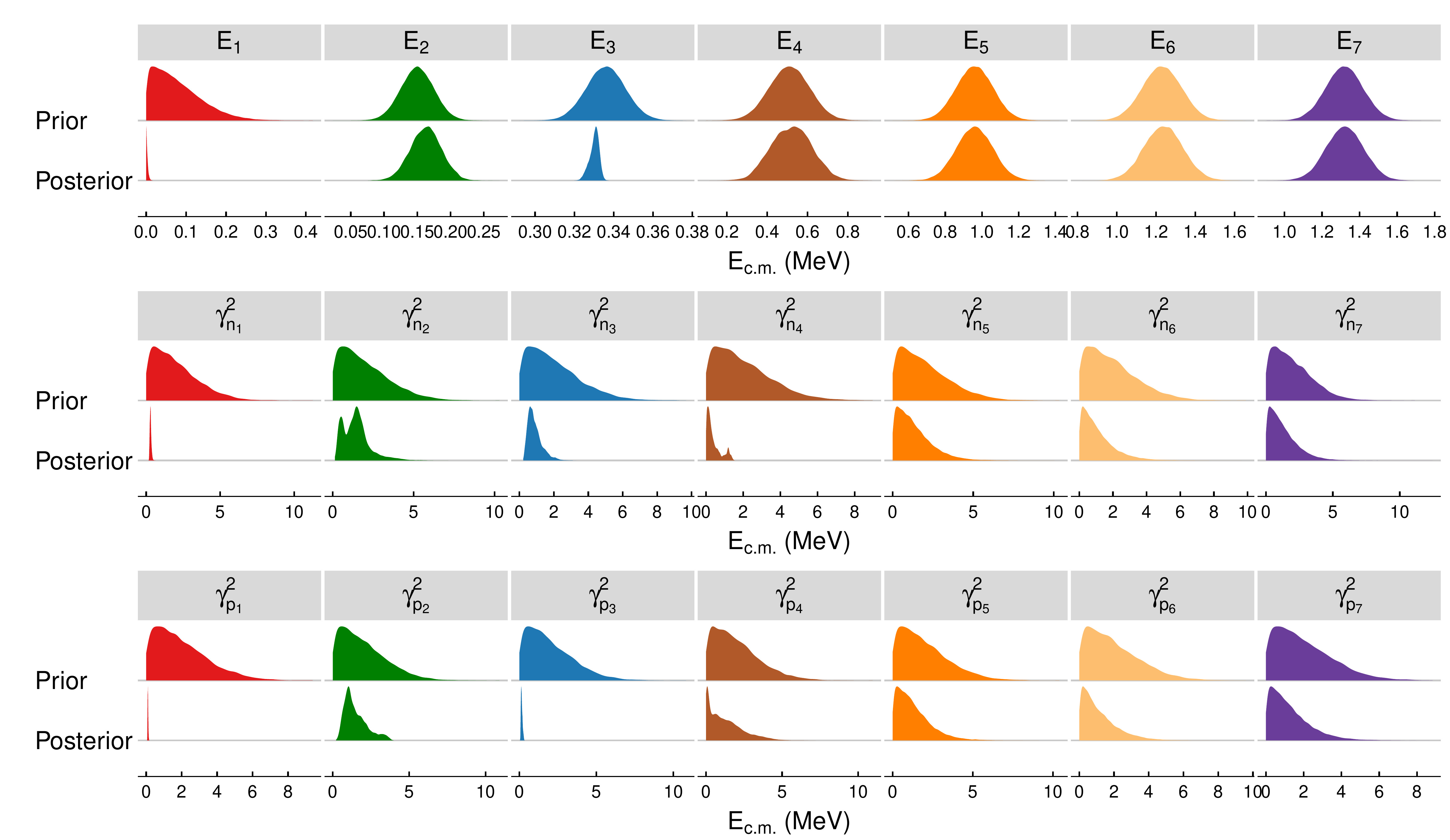}
\caption{Comparison of prior and posterior densities for the R-matrix parameters of the seven resonances used in our Bayesian model. New information from our fitting is only obtained for the first three resonances, for which the posterior is significantly narrower than the prior.}
\label{fig:Be7_Rmatrix}
\end{figure*}

Our predicted R-matrix parameters ($E_r$, $\gamma^2_n$, $\gamma^2_p$) for the first seven $^8$Be levels above the neutron threshold are listed in Table~\ref{tab:rmatrix}. For better comparison to the literature, we also list values for derived quantities ($E_x$, $\Gamma_n$, $\Gamma_p$, $\Gamma$). Figure~\ref{fig:Be7_Rmatrix} compares the prior and posterior densities for all R-matrix parameters. The values containing new information, i.e., with a posterior significantly different than the prior, are shown in boldface in Table~\ref{tab:rmatrix}. Before discussing individual levels, we note that our fit yields channel radii of $a_n$ = $a_p$ = $\pp{3.9}{0.5}$~fm. However, these values reflect the priors (Equation~(\ref{eq:model})) and thus no new information for the channel radii could be extracted from our fit.

For the 2$^-$ level, we find an excitation energy of $E_x$ $=$ $18900.6^{+2.8}_{-1.5}$~keV, just 2~keV above the neutron threshold. Recall that ENSDF lists this level without an energy uncertainty (Section~\ref{sec:2-}). However, since we assumed for the resonance energy a normal prior that is restricted to positive values only (Table~\ref{tab:nuclear}), our value for the energy is likely biased. In any case, small shifts in energy for this broad level should not impact the total cross section fit noticeably. Our prediction for the total width, $\Gamma$ $=$ $\Gamma_n$ $+$ $\Gamma_p$ $=$ $161^{+23}_{-19}$~keV, agrees with previous results but has a smaller uncertainty.

For the first 3$^+$ level, no new information is obtained for the excitation energy (Figure~\ref{fig:Be7_Rmatrix}). The total width amounts to $\Gamma_p$ $=$ $1.04\pm0.37$~MeV, which agrees with previous values within uncertainties. For the second 3$^+$ level, we find an excitation energy of $E_x$ $=$ $19229.1^{+2.0}_{-2.9}$~keV, which can be compared to the value of $19235\pm10$~keV from ENSDF (Table~\ref{tab:nuclear}). Our total width, $\Gamma_p$ $=$ $239\pm24$~keV, agrees with previous results. 

For the energies of the 1$^-$, 4$^+$, 2$^+$, and 0$^+$ levels, the posteriors are close to the priors (Figure~\ref{fig:Be7_Rmatrix}) and thus no new information could be derived from our fit. Similar arguments apply to the neutron and proton reduced widths. In addition, the posteriors of the reduced widths reveal a significant coverage probability near zero, and, therefore, only upper limits could be extracted. These are also listed in Table~\ref{tab:rmatrix}. Our upper limits for the total widths are consistent with previous values. The only exception is the 4$^+$ state, for which we obtain $\Gamma$ $<$ $86$~keV, which is smaller than the literature values (Table~\ref{tab:nuclear}). However, a direct comparison is not straightforward, considering that our fit is not very sensitive to the properties of this state and the previous study has reported too optimistic uncertainties, as discussed in Section~\ref{sec:4+}.

\begin{deluxetable*}{llllllll}
\tablecaption{R-matrix parameters obtained from our Hierarchical Bayesian Model\tablenotemark{a}. The quoted uncertainties correspond to 68\% credible intervals around the median value. Upper limits refer to 95th percentiles. New information is shown in boldface.\label{tab:rmatrix}} 
\tablewidth{2\columnwidth}
\tablehead{
$J^{\pi}$  &  $E_r$ (MeV)                        &   $\gamma_n^2$ (MeV)                 &  $\gamma_p^2$ (MeV)                   &  E$_x$ (MeV)\tablenotemark{b}               &  $\Gamma_n$ (MeV)\tablenotemark{c}      & $\Gamma_p$ (MeV)\tablenotemark{c}     & $\Gamma$ (MeV)\tablenotemark{c} 
} 
\startdata
$2^-$   & $\mathbf{\qq{0.0020}{0.0015}{0.0029}}$ & $\mathbf{\qq{0.291}{0.045}{0.054}}$  & $\mathbf{\pp{0.097}{0.020}}$          & $\mathbf{\qq{18.9006}{0.0015}{0.0028}}$     & $\mathbf{\pp{0.020}{0.011}}$            & $\mathbf{\qq{0.140}{0.013}{0.017}}$   & $\mathbf{\qq{0.161}{0.019}{0.023}}$   \\
$3^+$   & $\pp{0.162}{0.023}$                    & $\mathbf{\qq{1.38}{0.81}{0.69}}$     & $\mathbf{\qq{1.29}{0.48}{0.94}}$      & $\pp{19.061}{0.023}$                        & $\mathbf{\qq{0.071}{0.029}{0.037}}$     & $\mathbf{\pp{0.97}{0.33}}$            & $\mathbf{\pp{1.04}{0.37}}$            \\
$3^+$   & $\mathbf{\qq{0.3305}{0.0029}{0.0020}}$ & $\mathbf{\qq{0.80}{0.28}{0.48}}$     & $\mathbf{\qq{0.139}{0.041}{0.061}}$   & $\mathbf{\qq{19.2291}{0.0029}{0.0020}}$     & $\mathbf{\qq{0.131}{0.038}{0.025}}$     & $\mathbf{\qq{0.115}{0.020}{0.016}}$   & $\mathbf{\pp{0.239}{0.024}}$          \\
$1^-$   & $\pp{0.52}{0.10}$                      & $< 1.2$                              & $< 3.5$                               & $\pp{19.42}{0.10}$                          & $< 1.4$                                 & $< 5.8$                               & $< 6.3$                               \\
$4^+$   & $\pp{0.96}{0.10}$                      & $< 3.5$                              & $< 3.4$                               & $\pp{19.86}{0.10}$                          & $< 0.0062$                              & $< 0.083$                             & $< 0.086$                             \\
$2^+$   & $\pp{1.24}{0.10}$                      & $< 3.1$                              & $< 3.2$                               & $\pp{20.14}{0.10}$                          & $< 2.2$                                 & $< 3.9$                               & $< 4.8$                               \\
$0^+$   & $\pp{1.32}{0.10}$                      & $< 3.4$                              & $< 3.4$                               & $\pp{20.22}{0.10}$                          & $< 2.5$                                 & $< 4.1$                               & $< 5.5$                               \\
\enddata
\tablenotetext{a}{Our fit resulted in channel radii of $a_n$ = $a_p$ = $\pp{3.9}{0.5}$~fm, which mainly reflect the chosen priors (see text).}
\tablenotetext{b}{Excitation energies are calculated from the resonance energy (column 2) and $S_n$ $=$ 18898.64$\pm$0.08~keV \citep{wan17}.}
\tablenotetext{c}{Values are derived from the posterior distributions of the partial and total widths, which are calculated from Equation~(\ref{eq:gamwidth}), and thus take all parameter correlations into account.} 
\end{deluxetable*}

\begin{figure}[hbt!]
\includegraphics[width=1\linewidth]{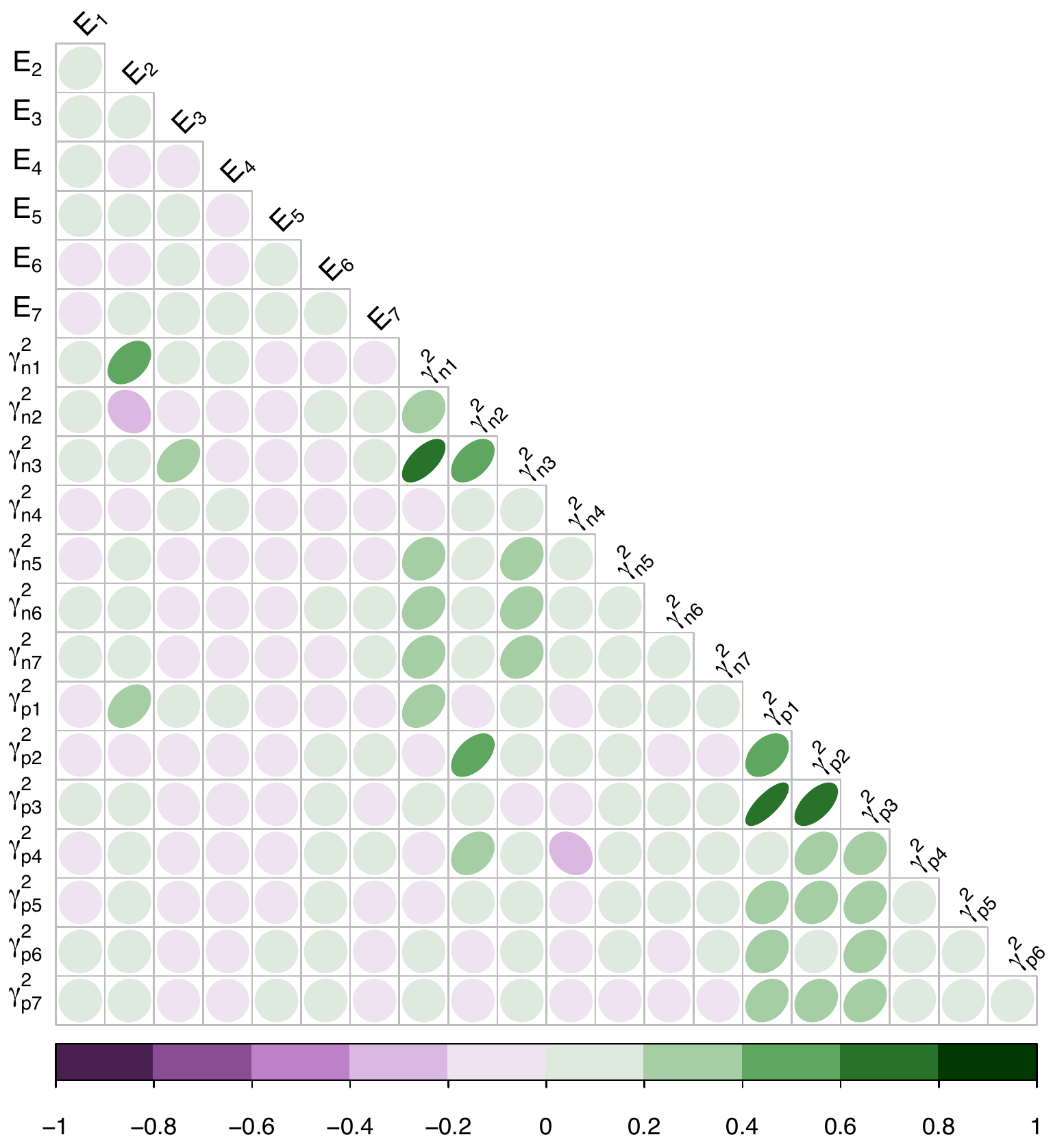}
\caption{ Pair-wise correlations of the R-matrix parameters ($E_0$, $\gamma_n^2$, $\gamma_p^2$) for each of the seven resonances, labeled $1$ to $7$. The purple and green colors represent a negative and positive correlation, respectively. The eccentricity of the ellipses is a measure for the magnitude of the correlation.}
\label{fig:Be7_correlation}
\end{figure}

 Figure~\ref{fig:Be7_correlation} shows the pair-wise correlations of the fitted R-matrix parameters for each of the seven resonances. Negative and positive correlations are depicted in purple and green, respectively. The eccentricity of the ellipses is proportional to the strength of the correlation. The strongest correlations are apparent for the reduced widths of the first three resonances.

\section{Reaction rates of \BeLi}\label{sec:rates}
The thermonuclear reaction rate per particle pair, $N_A \langle \sigma v \rangle$, at a given plasma temperature, $T$, is given by \citep{Iliadis:2015ta}
\begin{align}
N_A \langle \sigma v \rangle 
 = \sqrt{\frac{8}{\pi m_{01}}} \frac{N_A}{(kT)^{3/2}} 
    \int_0^\infty \sqrt{E}\sigma_{np}(E)\,\sqrt{E}\,e^{-E/kT}\,dE, 
\label{eq:rate}
\end{align}
where $m_{01}$ is the reduced mass of projectile and target, $N_A$ is  Avogadro's constant, $k$ is the Boltzmann constant, and $E$ is the $^7$Be $+$ $n$ center-of-mass energy.

We computed the \BeLi reaction rates by numerically integrating Equation~(\ref{eq:rate}). The reduced cross section is calculated from the samples of the Bayesian R-matrix fit, discussed in Section~\ref{sec:redcross}, and thus our values of $N_A \langle \sigma v \rangle$ fully contain the effects of statistical, systematic, and extrinsic uncertainties, and of varying channel radii. We base these results on 10,000 random reduced cross section samples, which ensures that Markov chain Monte Carlo fluctuations are negligible compared to the reaction rate uncertainties. Our lower integration limit was set at $10^{-4}$~eV. Reaction rates are computed for $46$ different temperatures between $1$~MK and $1$~GK. At these temperatures, the data shown in Figure~\ref{fig:Be7_Rmatrix} fully cover the astrophysically important energy range. Numerical values of the reaction rates are listed in Table~\ref{tab:rate}. The recommended rates are computed from the 50th percentile of the probability density, while the factor uncertainty, $f.u.$, is obtained from the 16th and 84th percentiles \citep{Longland:2010is}. Our total rate uncertainties range from 1.5\% to 2.1\% for the entire temperature range shown.
\begin{deluxetable}{ccc|ccc}
\tablecaption{Recommended $^7$Be(n,p)$^7$Li reaction rates.\tablenotemark{a}
\label{tab:rate}} 
\tablewidth{\columnwidth}
\tabletypesize{\footnotesize}
\tablecolumns{7}
\tablehead{
  T (GK) & Rate   & $f.u.$ & T (GK) & Rate  & $f.u.$} 
\startdata
0.001 & 5.157E+09 & 1.016 &   0.070 & 3.457E+09 & 1.017 \\ 
0.002 & 5.025E+09 & 1.016 &   0.080 & 3.362E+09 & 1.017 \\ 
0.003 & 4.928E+09 & 1.015 &   0.090 & 3.278E+09 & 1.017 \\ 
0.004 & 4.849E+09 & 1.015 &   0.100 & 3.200E+09 & 1.017 \\ 
0.005 & 4.781E+09 & 1.015 &   0.110 & 3.128E+09 & 1.018 \\ 
0.006 & 4.722E+09 & 1.015 &   0.120 & 3.063E+09 & 1.018 \\ 
0.007 & 4.667E+09 & 1.015 &   0.130 & 3.002E+09 & 1.018 \\ 
0.008 & 4.618E+09 & 1.015 &   0.140 & 2.944E+09 & 1.018 \\ 
0.009 & 4.573E+09 & 1.015 &   0.150 & 2.891E+09 & 1.018 \\ 
0.010 & 4.531E+09 & 1.015 &   0.160 & 2.840E+09 & 1.018 \\ 
0.011 & 4.491E+09 & 1.015 &   0.180 & 2.747E+09 & 1.019 \\ 
0.012 & 4.454E+09 & 1.015 &   0.200 & 2.663E+09 & 1.019 \\ 
0.013 & 4.419E+09 & 1.015 &   0.250 & 2.486E+09 & 1.019 \\ 
0.014 & 4.386E+09 & 1.015 &   0.300 & 2.343E+09 & 1.019 \\ 
0.015 & 4.354E+09 & 1.015 &   0.350 & 2.225E+09 & 1.020 \\ 
0.016 & 4.323E+09 & 1.015 &   0.400 & 2.127E+09 & 1.020 \\ 
0.018 & 4.266E+09 & 1.015 &   0.450 & 2.043E+09 & 1.020 \\ 
0.020 & 4.213E+09 & 1.015 &   0.500 & 1.971E+09 & 1.020 \\ 
0.025 & 4.096E+09 & 1.015 &   0.600 & 1.857E+09 & 1.020 \\ 
0.030 & 3.994E+09 & 1.016 &   0.700 & 1.771E+09 & 1.020 \\ 
0.040 & 3.823E+09 & 1.016 &   0.800 & 1.706E+09 & 1.020 \\ 
0.050 & 3.682E+09 & 1.016 &   0.900 & 1.653E+09 & 1.021 \\ 
0.060 & 3.562E+09 & 1.016 &   1.000 & 1.612E+09 & 1.021 \\ 
\enddata
\tablenotetext{a}{In units of cm$^3$~mol$^{-1}$~s$^{-1}$, corresponding to the 50th percentiles of the rate probability density function. The rate factor uncertainty, $f.u.$, corresponds to a coverage probability of 68\% and is obtained from the 16th and 84th percentiles.}
\end{deluxetable}

\begin{figure}[hbt!]
\includegraphics[width=1\linewidth]{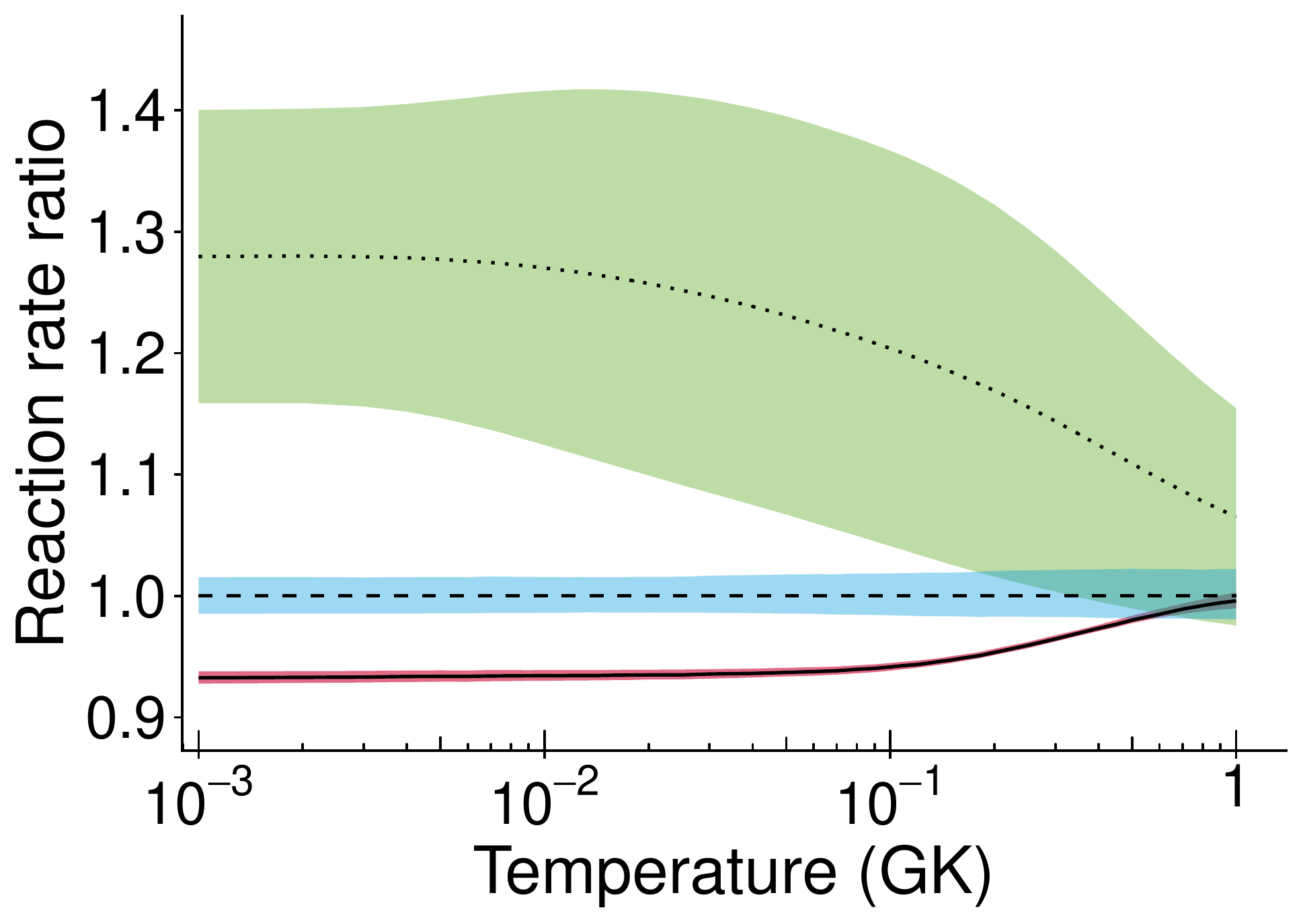}
\caption{Present $^7$Be(n,p)$^7$Li thermonuclear reaction rates (cyan), compared to the evaluations of \citealt{des04} (magenta) and \citealt{dam18} (green). Our results correspond to 68\% coverage probabilities, while the previous uncertainties have been obtained from $\chi^2$ analyses. All rates are normalized to our new recommended rate, listed in Table~\ref{tab:rate}. The most important temperature for $^7$Be and $^7$Li big bang nucleosynthesis is near 0.5~GK.}
\label{fig:ratecomp}
\end{figure}

Reaction rates are displayed in Figure~\ref{fig:ratecomp}. Our low (16th percentile) and high (84th percentile) rates are shown as a cyan band centered around unity. The results from the previous evaluations of \citet{des04}, which are widely used in big bang nucleosynthesis simulations \citep[see][and references therein]{pit18}, and the recently published rates of \citet{dam18} are displayed as magenta and green bands, respectively. The uncertainties of the previous rates were obtained from $\chi^2$ analyses and have a different meaning compared to the present work. All rates shown are normalized to the present median rates (50th percentile). 

It is apparent that the rates of \citet{des04} are lower over almost the entire temperature range shown and have also much smaller uncertainties. Near a temperature of $T$ $=$ $0.5$~GK, which is most important for the fig bang nucleosynthesis of $^7$Be and $^7$Li, the present and previous rates agree marginally within uncertainties. However, our rate uncertainty is 2.0\%, while \citet{des04} report a value of only 0.6\%. As can be seen from Graph~1j in \citet{des04}, their reaction rates are determined by the absolute cross section normalization of the \citet{koe88} data. Since the systematic and statistical uncertainties for the latter experiment amount to $\approx$2.0\% and $\approx$1.0\% (Table~\ref{tab:norm}), respectively, it is not clear why the rate uncertainties derived by \citet{des04} ended up to be much less than $1.0$\%. In comparison, our absolute reaction rate scale is determined by the reduced cross section normalizations of six independent measurements (Section~\ref{sec:bay7be} and Figure~\ref{fig:Be7_norm}), three of which were published after the evaluation of \citet{des04}. 

The rates of \citet{dam18} are higher than our results at most temperatures shown. Near 0.5~GK, the rates are in marginal agreement but the previous rate uncertainty is large and amounts to 12.5\% compared to 2.0\% for the present rates. In their analysis, \citet{dam18} used only their own data, which have a significantly larger magnitude compared to all other results (square symbols in Figures~\ref{fig:data} and \ref{fig:MCMCBe7}), and the data of \citet{sek76}, which we disregarded for reasons explained in Section~\ref{sec:disregardedExp}.

\section{Summary}
\label{sec:summary}
The goal of the present work was to derive new \BeLi thermonuclear reaction rates based on all available experimental information. This reaction sensitively impacts the primordial abundances of $^{7}$Be and $^7$Li during big bang nucleosynthesis. Of particular interest is not only an improved recommended rate, but also the estimation of reliable rate uncertainties. 

 We started by critically evaluating all available data and disregarding experimental results that were either questionable or provided insufficient information about statistical and systematic uncertainties. The reduced cross section data we adopted for our analysis reveal significant discrepancies. Some of the reduced cross sections measured by different groups differ by up to $\approx$30\%. Therefore, a robust derivation of thermonuclear reaction rates is challenging.

 We presented a solution to this problem by combining a hierarchical Bayesian model with an R-matrix model to analyze the remaining six data sets we deemed most reliable. The nuclear structure of $^8$Be near the neutron threshold has been evaluated to estimate appropriate prior densities for our analysis. In the fitting, we fully implemented statistical, systematic, and extrinsic uncertainties, and also took the variation of the neutron and proton channel radii into account. The fitting of the data was performed with an automated factor slice sampler using a Markov chain of length 500,000. 

From the posteriors, we extracted R-matrix parameters ($E_r$, $\gamma^2_n$, $\gamma^2_p$) and derived excitation energies, partial and total widths. Our fit is sensitive to the contributions of the first three levels (2$^-$, 3$^+$, 3$^+$) above the neutron threshold. Our values of $E_x$ and $\Gamma$ for these states are in overall agreement with previous results, but our results have significantly smaller uncertainties. 

Reaction rates were computed by integrating 10,000 credible samples of the reduced cross section. Our results are compared to the previously published rates of \citet{des04} and \citet{dam18}. The three rates are in marginal agreement at a temperature near 0.5~GK, which is most important for $^7$Be and $^7$Li big bang nucleosynthesis. At this temperature, our rate uncertainty amounts to 2.0\%, compared to 0.6\% for \citet{des04} and 12.5\% for \citet{dam18}. The previous evaluations only analyzed a subset of the data that were taken into account in the present work. 

 The uncertainty in our reduced cross section fit (a few percent) is much smaller than the discrepancy between some data sets ($\approx$30\%). One may then reasonably ask how reliable our results are considering the possibility that some authors may have underpredicted their reported systematic uncertainties. While we cannot exclude this possibility, we can claim that our fitted reduced cross section and derived thermonuclear reaction rate provide the best quantitative estimate available at this time based on all of the available experimental information.

\acknowledgments
We would like to thank Caleb Marshall and Richard Longland for providing constructive feedback, Guido Mart\'in-Hern\'andez for providing us with the original data,  and John Kelley for his help with evaluating the $^8$Be level properties. This work was supported in part by NASA under the Astrophysics Theory Program grant 14-ATP14-0007 and by U.S. DOE under contracts DE-FG02-97ER41041 (UNC) and DE-FG02-97ER41033 (TUNL).

\appendix
\section{Nuclear Data for the \BeLi Reaction}
\label{sec:app}
\subsection{Masses, energies, and reciprocity}
\label{sec:reciprocity}
\twocolumngrid

The data analyzed in this work are either adopted from $^7$Be(n,p)$^7$Li experiments or from measurements of the reverse reaction, $^7$Li(p,n)$^7$Be. 

The adopted masses and energies are listed in Table~\ref{tab:massenergy}. Atomic and nuclear masses are related by
\begin{equation}
m_{at}(A,Z) = m_{nu}(A,Z) + Z m_e - B_e(Z),
\end{equation}
where $A$ and $Z$ denote the mass number and atomic number, respectively; $m_e$ is the electron rest mass, and $B_e(Z)$ is the total electron binding energy in the neutral atom of atomic number $Z$. A positive sign is assigned to the binding energy, $B_e$. The Q-values based on nuclear and atomic masses are related by \citep[see also][]{ili19}
\begin{equation}
\label{eq:qan}
Q_{nu} = Q_{at} + \left( \Sigma B_e^i - \Sigma B_e^f \right),
\end{equation}
where $\Sigma B_e^i$ and $\Sigma B_e^f$ are the sum of total electron binding energies before and after the interaction, respectively. From the masses and energies listed in Table~\ref{tab:massenergy}, we find for the $^7$Be(n,p)$^7$Li reaction values of $Q_{nu}$ $=$ $1644.425\pm0.075$~keV and $Q_{at}$ $=$ $1644.229\pm0.075$~keV, where the main source of uncertainty derives from the $^7$Be atomic mass. Notice the $\approx$ $0.2$~keV difference as a result of taking the electron binding energies into account. 

\begin{deluxetable*}{llll}
\label{tab:massenergy} 
\tablecaption{Masses and energies adopted in the present work.}
\tablewidth{\linewidth}
\tablehead{
Quantity  &  Symbol &  Value(Uncertainty)  & Reference 
} 
\startdata
   Atomic mass unit energy          &      m$_uc^2$  &  931.4940954(57)~MeV                &     \citet{moh16}  \\ 
   Electron mass                    &      m$_e$     &   0.000548579909067(14)(9)(2)~u     &     \citet{stu14}   \\
   Proton mass                      &      m$_p$     &   1.007276466583(15)(29)~u          &     \citet{hei17}   \\
   Neutron mass                     &      m$_n$     &   1.008664915823(491)~u             &     \citet{wan17}   \\
   Atomic $^7$Li mass               &      m($^7$Li) &  7.016003437(5)~u                   &     \citet{wan17}   \\
   Atomic $^7$Be mass               &      m($^7$Be) &  7.01692872(8)~u                    &     \citet{wan17}   \\
   Electron binding energy of Li    &      B$_e$(Li) &  203.486169(2)~eV \tablenotemark{a} &     \citet{kra18}   \\
   Electron binding energy  of Be   &      B$_e$(Be) &  399.14864(5)~eV \tablenotemark{a}  &     \citet{kra18}  \\
\enddata
\tablenotetext{a}{See also \citet{hua76}.}
\end{deluxetable*}

Some of the $^7$Be(n,p)$^7$Li data analyzed in the present work were obtained by applying the reciprocity theorem to original $^7$Li(p,n)$^7$Be data. In our case, the theorem is given by \citep{bla52}
\begin{equation}
\frac{\sigma_{^7\!Be+n}}{\sigma_{^7\!Li+p}} = \frac{p_{^7\!Li+p}^2}{p_{^7\!Be+n}^2}, 
\end{equation}
where the linear momentum, $p$, is associated with the relative motion of the two particles in the laboratory system, or, equivalently, the motion of one of the particles in the center-of-mass system.

Assuming classical kinematics and {\it nuclear} masses, the reciprocity theorem can be written as
\begin{equation}
\frac{\sigma_{^7\!Be+n}}{\sigma_{^7\!Li+p}} = \frac{m_{^7\!Li+p}E_{^7\!Li+p}}{m_{^7\!Be+n}E_{^7\!Be+n}}, 
\end{equation}
with $m_{X+x}$ the reduced mass and $E_{X+x}$ the center-of-mass energy.  From the numerical values of the masses (Table~\ref{tab:massenergy}) and the $^7$Be(n,p)$^7$Li reaction Q-value derived from nuclear masses (Equation~(\ref{eq:qan})), we find numerically
\begin{equation}\label{eq:clas1}
\frac{\sigma_{^7\!Be+n}}{\sigma_{^7\!Li+p}} = \frac{0.99878958}{1 - 1.14360209\frac{Q_{np}}{E_p}}, 
\end{equation}
with $E_p$ the laboratory proton energy. The kinetic energy in the $^7$Be $+$ $n$ center-of-mass system is related to the laboratory proton kinetic energy in the $^7$Li $+$ $p$ system by

\begin{equation}\label{eq:clas2}
E_{^7\!Be+n} = 0.87443002 E_p - Q_{np}.
\end{equation}

Applying relativistic kinematics to the $^7$Li(p,n)$^7$Be reaction (see, e.g., \citet{byc72}), the invariant energy, $\sqrt{s}$, which is the same in all reference frames, can be written as 
\begin{equation}\label{eq:rel1}
s = m_p^2 + m_{^7\!Li}^2 + 2m_{^7\!Li}(E_p + m_p),
\end{equation}
where $E_p$ and $m_x$ denote the proton kinetic energy and the {\it nuclear} rest mass, respectively. All quantities are given in units of energy. The total kinetic energy in the $^7$Be $+$ p center-of-momentum system is 
\begin{equation}\label{eq:rel2}
E_{^7\!Be+n} = \sqrt{s} - m_{n} - m_{^7\!Be},
\end{equation}
and the center-of-momentum linear momentum of one particle in a given channel is given by
\begin{equation}\label{eq:rel3}
p_{^7\!Li+p} = \frac{\sqrt{\left[s-(m_p+m_{^7\!Li})^2\right]\left[s-(m_p-m_{^7\!Li})^2\right]}}{2\sqrt{s}},   \\
\end{equation}
\begin{equation}\label{eq:rel4}
p_{^7\!Be+n} = \frac{\sqrt{\left[s-(m_n+m_{^7\!Be})^2\right]\left[s-(m_n-m_{^7\!Be})^2\right]}}{2\sqrt{s}}.    
\end{equation}

\begin{figure}[hbt!]
\includegraphics[width=1\linewidth]{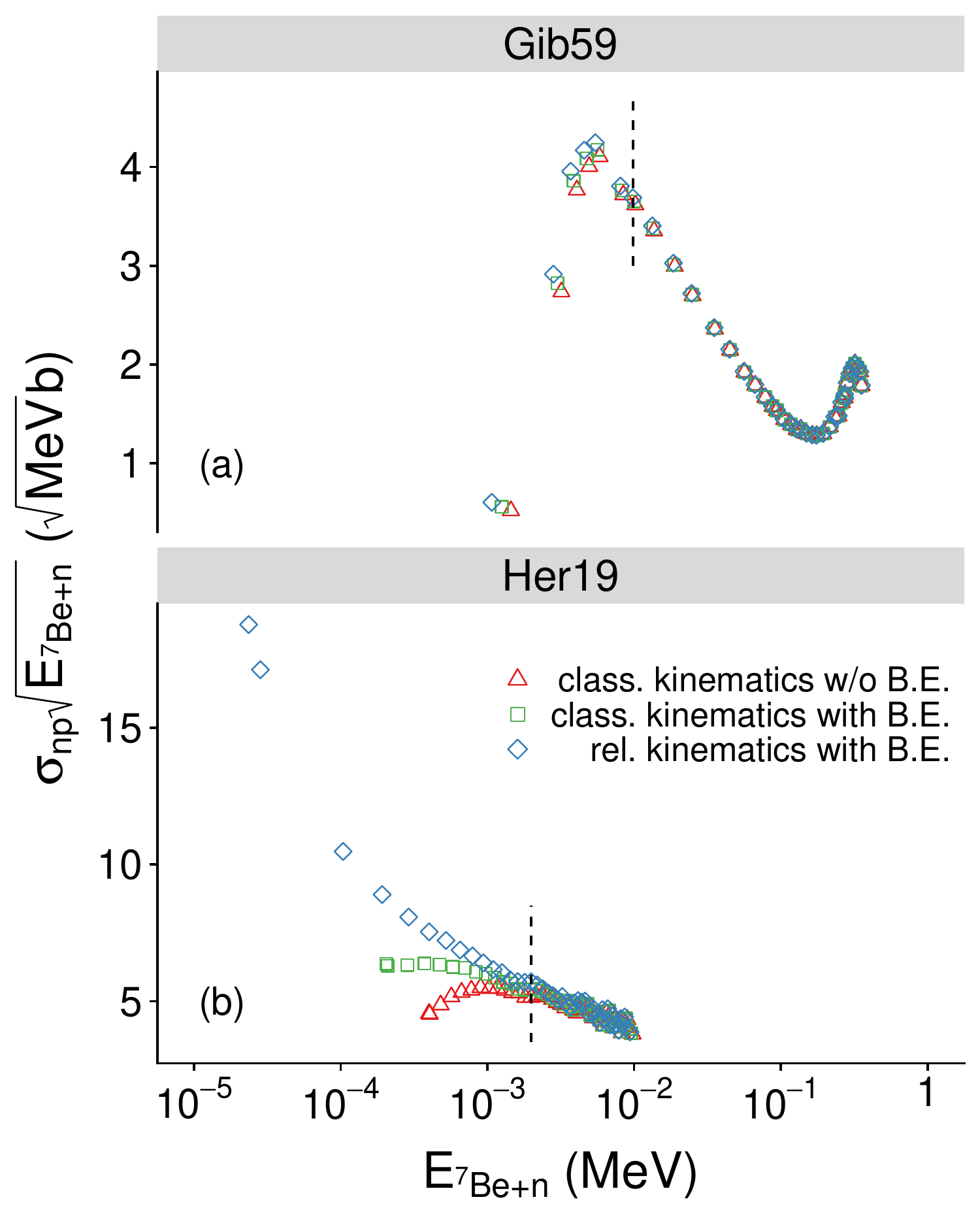}
\caption{\label{fig:compData} 
Reduced cross sections of $^7$Be(n,p)$^7$Li derived from $^7$Li(n,p)$^7$Be data using the reciprocity theorem. The abscissa shows the total kinetic energy in the $^7$Be+n center-of-momentum system. Top panel: Original (p,n) data from \citet{gib59}, and converted in the present work to (n,p) reduced cross sections using different assumptions: (red triangles) classical kinematics and atomic masses (i.e., electron binding energies are set to zero); (green squares) classical kinematics and nuclear masses (i.e., adopting electron binding energies); (blue diamonds) relativistic kinematics and nuclear masses. Bottom panel: Same for the original data of \citet{mar19}. All data on the low-energy side of the dashed vertical lines were disregarded in the present work.
}
\end{figure}

The impact of the different assumptions is demonstrated in Figure~\ref{fig:compData} for two $^7$Li(p,n)$^7$Be data sets \citep{gib59,mar19}. The abscissa and ordinate show the reduced (n,p) cross section, calculated using the reciprocity theorem, and the center-of-momentum kinetic energy in the $^7$Be+n system, respectively. Panel (a)  was obtained by transforming the $^7$Li(p,n)$^7$Be data of \citet{gib59}. The red triangles, green squares, and blue diamonds are obtained assuming classical kinematics and disregarding electron binding energies (Equations~(\ref{eq:clas1}) and (\ref{eq:clas2})), classical kinematics and proper electron binding energies (Table~\ref{tab:massenergy}), and relativistic kinematics and proper electron binding energies (Equations~(\ref{eq:rel1}) $-$ (\ref{eq:rel4})), respectively. Small differences are noticeable only at the lowest-energy data points. However, they are inconsequential for our purposes because we disregarded data points below an energy of $10$~keV (vertical dashed line) that are impacted by experimental artifacts (see Section~\ref{sec:gibbons}).

The situation is different for the transformed $^7$Li(p,n)$^7$Be data of \citet{mar19}, which are shown in  panel (b). This experiment recorded data much closer to the (p,n) threshold compared to the results shown in panel (a). Therefore, small changes in the masses or kinematic expressions will have a large impact. The reduced (n,p) cross sections for the different assumptions diverge strongly below an energy of $1$~keV. In addition, the uncertainty in the $^7$Be atomic mass introduces a systematic uncertainty in excess of 2\% below an energy of $2$~keV (dashed line). Therefore, we disregarded the data on the low-energy side of the vertical dashed line.

Finally, in our context, the reciprocity theorem only applies to transitions connecting the ground state of $^7$Be and the ground state of $^7$Li. For proton energies in excess of $E_p$ $=$ $2371$~keV, the $^7$Li(p,n)$^7$Be reaction can also populate the first excited state of $^7$Be at $429.1$~keV. Therefore, we disregarded all $^7$Li(p,n)$^7$Be data that were measured above this proton bombarding energy.

\subsection{The \BeLi data of \citet{dam18}}
\citet{dam18} measured the $^{7}$Be(n,p)$^{7}$Li reaction at CERN's n$\_$TOF facility. Their results are tabulated at \url{https://twiki.cern.ch/twiki/bin/view/NTOFPublic}. It appears that their fit excludes their data above $35$~keV neutron energy in the laboratory system, which are shown as purple points in Figure~2 of \citet{dam18}. We transformed their listed laboratory neutron energies below $35$~keV to center-of-mass energies, and their listed cross sections to reduced cross sections. Their statistical uncertainties in this energy range amount to $1$ $-$ $10$\%. Their systematic uncertainties, which originate from angular distribution effects, sample mass determination, flux normalization, and detection efficiency estimation, are about $10$\%. \citet{dam18} also reported the cross section at thermal neutron energy. They found a value of 52300$\pm$5200~b. The reported uncertainty is dominated by systematic effects. For the statistical uncertainty, we assumed a value of 1\%.
The derived experimental cross sections, together with their statistical uncertainties are listed in Table~\ref{tab:damone}.

\begin{deluxetable}{cc|cc}

\tablecaption{The $^7$Be(n,p)$^7$Li data of \citet{dam18}.\label{tab:damone}} 
\tablewidth{\columnwidth}
\tablehead{
$E_{c.m.}$\tablenotemark{a}  & $\sqrt{E_{c.m.}}\sigma_{np}$\tablenotemark{a}  & $E_{c.m.}$\tablenotemark{a}   & $\sqrt{E_{c.m.}}\sigma_{np}$\tablenotemark{a}  \\
(MeV)  &  ($\sqrt{\mathrm{MeV}}$b) &  (MeV)  &  ($\sqrt{\mathrm{MeV}}$b) 
} 
\startdata
   1.7909E-08   &      7.809$\pm$0.027  &     2.8383E-05   &      7.73$\pm$0.14 \\
   2.8383E-08   &      7.819$\pm$0.016  &     4.4985E-05   &      7.67$\pm$0.16 \\
   4.4985E-08   &      7.780$\pm$0.015  &     7.1296E-05   &      7.52$\pm$0.17 \\
   7.1296E-08   &      7.766$\pm$0.017  &     1.1300E-04   &      7.83$\pm$0.20 \\
   1.1300E-07   &      7.652$\pm$0.024  &     1.7909E-04   &      7.51$\pm$0.21 \\
   1.7909E-07   &      7.571$\pm$0.037  &     2.8383E-04   &      7.24$\pm$0.25 \\
   2.8383E-07   &      7.573$\pm$0.046  &     4.4985E-04   &      6.97$\pm$0.26 \\
   4.4985E-07   &      7.617$\pm$0.052  &     7.1296E-04   &      6.83$\pm$0.28 \\
   7.1296E-07   &      7.571$\pm$0.059  &     1.1300E-03   &      7.00$\pm$0.31 \\
   1.1300E-06   &      7.498$\pm$0.066  &     1.7909E-03   &      5.65$\pm$0.30 \\
   1.7909E-06   &      7.574$\pm$0.075  &     2.8383E-03   &      5.40$\pm$0.33 \\
   2.8383E-06   &      7.576$\pm$0.084  &     4.4985E-03   &      4.98$\pm$0.34 \\
   4.4985E-06   &      7.667$\pm$0.094  &     7.1296E-03   &      5.15$\pm$0.39 \\
   7.1296E-06   &      7.51$\pm$0.10      &     1.1300E-02   &      3.55$\pm$0.32 \\
   1.1300E-05   &      7.74$\pm$0.12      &     1.7909E-02   &      3.49$\pm$0.33 \\
   1.7909E-05   &      7.41$\pm$0.13      &     2.8383E-02   &      3.83$\pm$0.42 \\
\enddata
\tablenotetext{a}{The original data are shown as black points in Figure 2 of \citet{dam18} and were provided to us by the authors. We converted their laboratory energies to center-of-mass energies, and their cross sections to reduced cross sections. We only list statistical uncertainties in this table. The systematic uncertainty is about 10\% (see text). Notice that the quoted results include the contributions for the $^7$Li ground and first excited states.}
\end{deluxetable}
%

\subsection{The \LiBe data of \citet{gib59}}
\label{sec:gibbons}
\citet{gib59} measured the cross section of the $^{7}$Li(p,n)$^{7}$Be reaction using a lithium metal target ($2.387$~MeV $\le$ $E_p$ $\le$ $5.418$~MeV, or $0.443$~MeV $\le$ $E_{^7Be+n}$ $\le$ $3.09$~MeV) and LiF targets ($1.882$~MeV $\le$ $E_p$ $\le$ $2.450$~MeV, or $1.44$~keV $\le$ $E_{^7Be+n}$ $\le$ $0.498$~MeV). They provide little detail regarding their measurement, but more information is given in their earlier work \citep{mac58} that investigated the energy region between the threshold and the resonance at $2.25$~MeV. We adopted the results shown in Figure~4 of \citet{gib59}, which those authors supplied to the compilation of \citet{kim65}, and converted the listed laboratory proton energies and (p,n) cross sections to center-of-mass neutron energies and reduced neutron cross sections, respectively (see Table~\ref{tab:gibbonsdata}). 
In addition, we disregarded all data points below $E_{^7Be+n}$ $=$ $10$~keV (see the vertical dashed line in part~(a) of Figure~\ref{fig:compData}), since these are subject to {\it ``the effect of a range of possible choices of effective target thicknesses.''}  The statistical uncertainties amount to $1$\%, as quoted in the captions of Figures~3 and 4 in \citet{gib60}. For the systematic uncertainty of the cross section normalization, we adopted the value of 5\% given in Figure~4 of \citet{gib59}.

\begin{deluxetable}{cc|cc}
\tablecaption{The transformed $^7$Be(n$_0$,p$_0$)$^7$Li data of \citet{gib59}.\label{tab:gibbonsdata}} 
\tablewidth{\columnwidth}
\tablehead{
$E_{c.m.}$\tablenotemark{a}  & $\sqrt{E_{c.m.}}\sigma_{np}$\tablenotemark{a}  & $E_{c.m.}$\tablenotemark{a}   & $\sqrt{E_{c.m.}}\sigma_{np}$\tablenotemark{a}  \\
(MeV)  &  ($\sqrt{\mathrm{MeV}}$b) &  (MeV)  &  ($\sqrt{\mathrm{MeV}}$b) 
} 
\startdata
9.8138E-03    &   3.688$\pm$0.037     &  1.9428E-01    &   1.300$\pm$0.013 \\
1.3311E-02    &   3.403$\pm$0.034     &  2.1526E-01    &   1.370$\pm$0.014 \\
1.8556E-02    &   3.025$\pm$0.030     &  2.3274E-01    &   1.470$\pm$0.015 \\
2.4676E-02    &   2.718$\pm$0.027     &  2.4673E-01    &   1.480$\pm$0.015 \\
3.5167E-02    &   2.372$\pm$0.024     &  2.5984E-01    &   1.616$\pm$0.016 \\
4.4783E-02    &   2.154$\pm$0.021     &  2.7208E-01    &   1.663$\pm$0.017 \\
5.6148E-02    &   1.929$\pm$0.019     &  2.7295E-01    &   1.698$\pm$0.017 \\
6.6639E-02    &   1.795$\pm$0.018     &  2.8257E-01    &   1.815$\pm$0.018 \\
7.8004E-02    &   1.670$\pm$0.017     &  2.9481E-01    &   1.906$\pm$0.019 \\
8.7620E-02    &   1.579$\pm$0.016     &  3.0792E-01    &   1.965$\pm$0.020 \\
9.3740E-02    &   1.537$\pm$0.015     &  3.2016E-01    &   2.009$\pm$0.020 \\
1.0510E-01    &   1.445$\pm$0.014     &  3.3240E-01    &   1.967$\pm$0.020 \\
1.1734E-01    &   1.393$\pm$0.014     &  3.4639E-01    &   1.927$\pm$0.019 \\
1.2783E-01    &   1.342$\pm$0.013     &  3.5513E-01    &   1.787$\pm$0.018 \\
1.3658E-01    &   1.348$\pm$0.013     &  3.7086E-01    &   1.703$\pm$0.017 \\
1.4969E-01    &   1.302$\pm$0.013     &  3.8310E-01    &   1.581$\pm$0.016 \\
1.6368E-01    &   1.286$\pm$0.013     &  3.9709E-01    &   1.505$\pm$0.015 \\
1.7504E-01    &   1.282$\pm$0.013     &  4.2069E-01    &   1.320$\pm$0.013\\
\enddata
\tablenotetext{a}{The original $^7$Li(p,n)$^7$Be data were adopted from \citet{kim65}. The $^7$Be(n,p)$^7$Li results are obtained from the application of the reciprocity theorem, the masses and energies listed in Table~\ref{tab:massenergy}, and relativistic kinematics (Equations~(\ref{eq:rel1}) $-$ (\ref{eq:rel4})). The statistical and systematic uncertainties amount to 1\% and 5\%, respectively (see text). We only list statistical uncertainties in this table. Notice that the quoted results include only the contribution of the $^7$Li ground state.}
\end{deluxetable}
%

%

%
 
 \subsection{The \LiBe data of \citet{mar16}}
 \citet{mar16} measured the shape of the $^{7}$Li(p,n)$^{7}$Be reaction cross section from threshold up to a proton laboratory energy of $1892$~keV. They state that their results {\it ``are normalized to Macklin and Gibbons and Newson data. Absolute determination of $\sigma_{pn}$ was not possible due to the lack of reliable proton beam current indication.''} A subsequent paper \citep{mar19} reports absolute cross sections from a thick target measurement, using a one-level Breit-Wigner formula and neglecting the $(E-E_0)^2$ term for $\ell$ $=$ $0$ captures following \citet{new57}. We adopted the data of \citet{mar16}, normalized to the results of \citet{mar19}; see the red data points in Figure~8 of the latter reference. The systematic uncertainty of the (p,n) cross section can be obtained from Equation~(10) of \citet{mar19}, together with their quoted value of ``$C_0$ $=$ $5.4$ $\pm$ $0.4$''. The systematic uncertainty ranges from 3.4\% to 7.7\% at the highest and lowest energies, respectively. In the present work, we adopt the geometric mean of 5.1\%. The data adopted in our analysis are listed in Table~\ref{tab:hernandezdata}.

\begin{deluxetable}{cc|cc}
\tablecaption{The transformed $^7$Be(n$_0$,p$_0$)$^7$Li data of \citet{mar19}.\label{tab:hernandezdata}} 
\tablewidth{\columnwidth}
\tablehead{
$E_{c.m.}$\tablenotemark{a}  & $\sqrt{E_{c.m.}}\sigma_{np}$\tablenotemark{a}  & $E_{c.m.}$\tablenotemark{a}   & $\sqrt{E_{c.m.}}\sigma_{np}$\tablenotemark{a}  \\
(MeV)  &  ($\sqrt{\mathrm{MeV}}$b) &  (MeV)  &  ($\sqrt{\mathrm{MeV}}$b) 
} 
\startdata
1.9874E-03   &  5.711$\pm$0.082    &    5.6034E-03   &  4.42$\pm$0.11  \\
2.1840E-03   &  5.595$\pm$0.084    &    5.8512E-03   &  4.22$\pm$0.11  \\
2.3859E-03   &  5.457$\pm$0.084    &    6.1003E-03   &  4.69$\pm$0.13  \\
2.5927E-03   &  5.306$\pm$0.084    &    6.3505E-03   &  4.35$\pm$0.12  \\
2.8044E-03   &  5.207$\pm$0.086    &    6.6016E-03   &  4.73$\pm$0.13  \\
3.0203E-03   &  5.052$\pm$0.085    &    6.8535E-03   &  4.14$\pm$0.12  \\
3.2404E-03   &  5.175$\pm$0.090    &    7.1059E-03   &  4.45$\pm$0.14  \\
3.4639E-03   &  4.925$\pm$0.090    &    7.3591E-03   &  4.13$\pm$0.13  \\
3.6909E-03   &  4.805$\pm$0.090    &    7.6127E-03   &  4.16$\pm$0.14  \\
3.9212E-03   &  4.939$\pm$0.094    &    7.8665E-03   &  3.94$\pm$0.15  \\
4.1544E-03   &  5.024$\pm$0.097    &    8.1203E-03   &  4.36$\pm$0.16  \\
4.3905E-03   &  4.971$\pm$0.098    &    8.3746E-03   &  4.14$\pm$0.15  \\
4.6289E-03   &  5.01$\pm$0.10        &    8.6291E-03   &  4.42$\pm$0.19  \\
4.8697E-03   &  4.567$\pm$0.099    &    8.8827E-03   &  4.02$\pm$0.20  \\
5.1127E-03   &  4.61$\pm$0.10        &    9.3896E-03   &  3.88$\pm$0.20  \\
5.3573E-03   &  4.69$\pm$0.11        &                          &        \\
\enddata
\tablenotetext{a}{The original $^7$Li(p,n)$^7$Be data are shown in Figure~8 of \citet{mar19} and were provided to us by the first author. The $^7$Be(n,p)$^7$Li results are obtained from the application of the reciprocity theorem, the masses and energies listed in Table~\ref{tab:massenergy}, and relativistic kinematics (Equations~(\ref{eq:rel1}) $-$ (\ref{eq:rel4})). The systematic uncertainty amounts to $\approx$5\% (see text). We only list statistical uncertainties in this table. Notice that the quoted results include only the contribution for the $^7$Li ground state.}
\end{deluxetable}

\subsection{The \BeLi data of \citet{koe88}}\label{sec:koehlerdata}
\citet{koe88} measured the $^7$Be(n,p)$^7$Li reaction at the Los Alamos Neutron Scattering Center (LANSCE) in the laboratory energy range from $0.025$~eV to $13.5$~keV. Their data are available in \citet{otu14}. We transformed the reported laboratory energies and cross sections to center-of-mass energies and reduced cross sections (see Table~\ref{tab:koehlerdata}). For energies below $E_{c.m.}$ $=$ 10$^{-4}$~MeV, the reduced cross section is constant and we binned the original data by grouping five adjacent data points and using the weighted average for the reduced cross section. This procedure reduces the large size of the data set without a significant loss in information. 
 
\citet{koe88} normalized their data to the $^7$Be(n,p)$^7$Li thermal neutron cross section, which they measured at the Omega West Reactor. Their value of the thermal cross section for the population of the $^7$Li ground state is $\sigma_{np_0}^{therm}$ $=$ 38,400$\pm$800~b (Table~\ref{tab:norm}). For the ratio of thermal cross sections for the first excited state and the ground state they find 0.011$\pm$0.003. The uncertainty of 2.1\% for the ground-state thermal cross section originates from the determination of the neutron flux, the number of $^7$Be target nuclei, the solid angle, and counting statistics, and, therefore, includes both statistical and systematic contributions. Notice that the ``relative'' uncertainties of the LANSCE data listed in \citet{otu14} also include a systematic component, because they include uncertainties for the cross section and anisotropy of the $^6$Li(n,$\alpha$)$^3$H monitor reaction. Without additional information it is not straightforward to disentangle the systematic and statistical uncertainties. Therefore, we adopted for the statistical uncertainties of the LANSCE data the ``relative'' uncertainties reported in \citet{otu14}. We also attribute the reported uncertainty of the thermal cross section (2.1\%) entirely to systematic effects. Furthermore, we assumed an uncertainty of 1\% for the statistical uncertainty of the thermal cross section.

\begin{deluxetable}{cc|cc}
\tablecaption{The $^7$Be(n,p)$^7$Li data of \citet{koe88}.\label{tab:koehlerdata}} 
\tablewidth{\columnwidth}
\tablehead{
$E_{c.m.}$\tablenotemark{a}  & $\sqrt{E_{c.m.}}\sigma_{np}$\tablenotemark{a}  & $E_{c.m.}$\tablenotemark{a}   & $\sqrt{E_{c.m.}}\sigma_{np}$\tablenotemark{a}  \\
(MeV)  &  ($\sqrt{\mathrm{MeV}}$b) &  (MeV)  &  ($\sqrt{\mathrm{MeV}}$b) 
} 
\startdata
   2.6088E-08  & 5.88 $\pm$0.13          &   1.5335E-05  & 5.63 $\pm$0.16  \\
   3.4569E-08  & 5.757$\pm$0.098         &   1.8290E-05  & 5.68 $\pm$0.16  \\
   4.7841E-08  & 5.68 $\pm$0.12          &   2.2242E-05  & 5.75 $\pm$0.17  \\
   6.1199E-08  & 5.720 $\pm$0.054        &   2.7575E-05  & 5.69 $\pm$0.16  \\
   7.3212E-08  & 5.695 $\pm$0.054        &   3.5094E-05  & 5.61 $\pm$0.16  \\
   8.9247E-08  & 5.733 $\pm$0.054        &   4.6197E-05  & 5.71 $\pm$0.18  \\
   1.1086E-07  & 5.664 $\pm$0.053        &   6.3560E-05  & 5.67 $\pm$0.18  \\
   1.4163E-07  & 5.627 $\pm$0.053        &   9.2796E-05  & 5.43 $\pm$0.16   \\
   1.8727E-07  & 5.674 $\pm$0.053        &   1.1978E-04  & 5.40 $\pm$0.36  \\
   2.5529E-07  & 5.712 $\pm$0.054        &   1.3202E-04  & 5.58 $\pm$0.31   \\
   3.1684E-07  & 5.711 $\pm$0.054        &   1.4601E-04  & 5.51 $\pm$0.29   \\
   3.8888E-07  & 5.800 $\pm$0.054        &   1.6262E-04  & 5.35 $\pm$0.29   \\ 
   4.8872E-07  & 5.654 $\pm$0.053        &   1.8273E-04  & 5.41 $\pm$0.31  \\ 
   6.3316E-07  & 5.752 $\pm$0.055        &   2.0546E-04  & 5.46 $\pm$0.31  \\
   8.5033E-07  & 5.820 $\pm$0.055        &   2.3431E-04  & 5.34 $\pm$0.31   \\
   1.2065E-06  & 5.759 $\pm$0.055        &   2.6841E-04  & 5.49 $\pm$0.36   \\
   1.5510E-06  & 5.691 $\pm$0.054        &   3.1037E-04  & 5.27 $\pm$0.33  \\
   1.7346E-06  & 5.839 $\pm$0.056        &   3.6370E-04  & 5.24 $\pm$0.34  \\
   1.9496E-06  & 5.656 $\pm$0.054        &   4.3277E-04  & 5.20 $\pm$0.33   \\
   2.2084E-06  & 5.708 $\pm$0.055        &   5.2195E-04  & 5.30 $\pm$0.34  \\
   2.5214E-06  & 5.676 $\pm$0.056        &   6.4259E-04  & 5.14 $\pm$0.30  \\
   2.9079E-06  & 5.623 $\pm$0.071        &   8.1046E-04  & 4.92 $\pm$0.28   \\
   3.3905E-06  & 5.676 $\pm$0.073        &   1.0579E-03  & 4.90 $\pm$0.27  \\
   4.0025E-06  & 5.655 $\pm$0.072        &   1.4251E-03  & 4.79 $\pm$0.25   \\
   4.7946E-06  & 5.662 $\pm$0.072        &   2.0371E-03  & 4.51 $\pm$0.22   \\
   5.8542E-06  & 5.660 $\pm$0.071        &   3.1387E-03  & 4.12 $\pm$0.20  \\
   7.3038E-06  & 5.657 $\pm$0.072        &   5.4730E-03  & 3.71 $\pm$0.15   \\
   9.3776E-06  & 5.762 $\pm$0.088        &   1.1803E-02  & 3.14 $\pm$0.22   \\
   1.2432E-05  & 5.723 $\pm$0.095        &   \\
\enddata
\tablenotetext{a}{The original data were retrieved from \citet{otu14}. Laboratory neutron energies and cross sections were transformed to center-of-mass energies and reduced cross sections. For energies below $E_{c.m.}$ $=$ 10$^{-4}$~MeV, where the reduced cross section is constant, we binned the original data by grouping five adjacent data points and using the weighted average for the reduced cross section. We only list statistical uncertainties in this table. The cross sections reported by \citet{koe88} were normalized using their measured thermal $^7$Be(n,p)$^7$Li cross section (see text), which has a systematic uncertainty of about 2\%. Notice that the quoted results include the contributions for the $^7$Li ground and first excited states.}
\end{deluxetable}

\subsection{Other thermal cross sections for \BeLi}
\citet{cer89} measured a total thermal cross section of 46800$\pm$4000~b at the VVR-S research reactor. From their reported branching ratio for populating the $^7$Li ground and first excited states, (2$\pm$1)\%, we find a thermal cross section of 45864$\pm$3972 for the ground state contribution (see Table~\ref{tab:norm}). The statistical uncertainty is 1\%. Systematic uncertainties originate from the number of $^7$Be target atoms (5\%), the number of $^6$Li atoms (5\%), the geometry, and possible variations of the neutron flux (5\%). 

The experiment of \citet{tom19} was carried out at the LVR-15 research reactor in the Czech Republic. They report a value of 43800$\pm$1400~b for the thermal ground state cross section. The systematic uncertainty originates from the $^6$Li thermal neutron cross section (0.4\%), the thickness of the $^6$LiF standard (0.7\%), the area of the standard (3.7\%), the $^7$Be activity (2\%), the distribution of $^7$Be atoms within the sample spot (0.8\%), and the geometry and beam profile (2.9\%). These values add quadratically to $5.2$\%, which exceeds the reported systematic uncertainty of $3.2$\%, indicating that the uncertainties are correlated. The statistical uncertainty is about 1.5\%, as is apparent from the results they quote for different measurements. 

\subsection{Disregarded experiments}\label{sec:disregardedExp}
We disregarded the $^{7}$Li(p,n)$^{7}$Be experiment of \citet{sek76}, although their data were used in previous analyses; see, e.g., \citet{ada03,dam18}. \citet{sek76} write {\it ``The discrepancy in measured yield in the region below $2.07$~MeV in fig. 4 between the polyethylene sphere data and the data of Macklin and Gibbons requires further discussion. The latter data are expected to be more nearly correct because of the expected weak dependence of the efficiency of the polyethylene sphere on (1) the neutron energy, for energies below $30$~keV and on (2) the angular distribution of the emitted neutrons. The data of fig. 4 give strong evidence that the combination of these two effects near the $^{7}$Li(p,n)$^{7}$Be threshold produces a change in efficiency of less than 7\% in a circumstance where the angular distribution change with proton bombarding energy is very rapid.''} It is apparent that they used the earlier results of \citet{gib59} to test the efficiency of their detector, and that they identified systematic problems with their own results that explain the $\approx$7\% deviation with respect to the work of \citet{gib59} at near-threshold energies. We conclude that their results, in terms of both absolute and relative cross sections, is of lower quality compared to the experiments of \citet{gib59} and \citet{mar19}. Note that the data of \citet{sek76} were also disregarded for similar reasons in the work of \citet{her14}.

The $^{7}$Be(n,p)$^{7}$Li measurement of \citet{and91} was performed at a single laboratory neutron energy of $24.5$~keV. They applied two normalization methods. The first was a cross section determination relative to the $^6$Li(n,$\alpha$)$^3$H reaction, which resulted in a rather large uncertainty of $\approx$30\%. The second was a normalization to the $^{7}$Be(n,p)$^{7}$Li thermal neutron cross section of \citet{koe88}. Only the weighted-average value with both methods ($\sigma_{np}$ $=$ 18$\pm$4 b) is reported by \citet{and91}, assuming isotropically emitted protons. Unfortunately, insufficient information is provided by the authors for disentangling the values resulting from their different methods, or for estimating separately the statistical and systematic uncertainties of their quoted cross section.
 
We also disregarded two measurements of the thermal $^7$Be(n,p)$^7$Li cross section: the  experiment by \citet{han55} found a value of 53000 $\pm$ 8000~b, which {\it ``is almost certainly too high due to the fact that he used too large a value for the branching ratio of the elctron capture of $^7$Be to the first excited state of $^7$Li,''}, as was pointed out by \citet{koe88}. The data of \citet{gle87} show a large scatter with large error bars, as can be seen from Figure~1 of \citet{dam18}. Their thermal neutron cross section also has a large uncertainty (50000$\pm$10000~b).

\begin{deluxetable*}{ccccccc}
\tablecaption{Thermal $^7$Be(n,p)$^7$Li cross sections and branching ratios.\tablenotemark{a}}
\label{tab:norm} 
\tablewidth{\columnwidth}
\tablehead{
$\sigma_{np_0}^{therm}$ \tablenotemark{b}  &  $\sigma_{np_1}^{therm}$/$\sigma_{np_0}^{therm}$ \tablenotemark{c}  &  $\sqrt{E_{c.m.}}\sigma_{np_0}^{therm}$  \tablenotemark{d}    &  Sys. \tablenotemark{e} & Stat. \tablenotemark{e} &   Reference \\
(b)    &   $\times$100  & ($\sqrt{\mathrm{MeV}}\mathrm{b}$)     &  (\%)    &     (\%) &     
} 
\startdata
43600$\pm$1600\tablenotemark{f} &            1.2$\pm$0.6    &    6.482    &    3.2\tablenotemark{f}            &    1.4\tablenotemark{f}                    &          \citet{tom19}          \\
52300$\pm$5200                  &                                    &    7.775    &    10            &    1.0\tablenotemark{g}   &          \citet{dam18}         \\
45864$\pm$3972                  &            2.0$\pm$1.0    &    6.818    &    8.5           &    1.0                              &          \citet{cer89}           \\
38400$\pm$800                   &            1.1$\pm$0.3    &    5.708    &    2.0           &    1.0\tablenotemark{g}   &          \citet{koe88}          \\
\enddata
\tablenotetext{a}{For a thermal neutron energy of $E_n$ $=$ $0.0253$~eV, corresponding to a neutron center-of-mass energy of $E_{c.m.}$ $=$ $0.0221$~eV.}
\tablenotetext{b}{Thermal $^7$Be(n,p)$^7$Li cross section for the transition to the $^7$Li ground state. The uncertainty includes statistical and systematic effects.}
\tablenotetext{c}{Ratio of thermal cross sections for populating the $^7$Li ground and first excited states.}
\tablenotetext{d}{Thermal $^7$Be(n,p)$^7$Li reduced cross section for the transition to the $^7$Li ground state.}
\tablenotetext{e}{Systematic and statistical uncertainties of the thermal cross section or thermal reduced cross section (columns 1 and 3, respectively).}
\tablenotetext{f}{Different values are quoted in the main text, abstract, and conclusions in \citet{tom19}. The values quoted here were obtained from the first author of that reference.}
\tablenotetext{g}{Assumed in the present work.}
\end{deluxetable*}

\bibliographystyle{aasjournal}
\bibliography{ref}
\end{document}